\documentclass[aps,prl,reprint,superscriptaddress,floatfix]{revtex4-2}

\usepackage[T1]{fontenc}
\usepackage{lmodern}
\usepackage{microtype}
\usepackage{amsmath,amssymb,mathtools,bm}
\usepackage{array,booktabs}
\usepackage[
    colorlinks= true,
    citecolor=blue
]{hyperref}
\usepackage{xcolor}
\newcommand{\ket}[1]{\lvert #1\rangle}
\newcommand{\bra}[1]{\langle #1\rvert}

\newcommand{\norm}[1]{\lVert #1\rVert}
\newcommand{\Tr}{\operatorname{Tr}}
\newcommand{\E}{\mathbb E}

\newcommand{\N}{\mathbb N}

\newcommand{\Haar}{\mathrm{Haar}}
\newcommand{\poly}{\operatorname{poly}}
\newcommand{\Sym}{\operatorname{Sym}}
\newcommand{\vct}[1]{\bm{#1}}
\newtheorem{theorem}{Theorem}
\definecolor{editred}{RGB}{200,0,0}

\begin{document}

\title{Classical Algorithms for Function Computation in Gaussian Boson Sampling}

\author{Ruoting Dou}
\affiliation{National Laboratory of Solid State Microstructures, School of physics and college of engineering and applied sciences, and Collaborative Innovation Center of Advanced Microstructures, Nanjing University, Nanjing 210093, China}

\author{Hao Zhan}
\affiliation{National Laboratory of Solid State Microstructures, School of physics and college of engineering and applied sciences, and Collaborative Innovation Center of Advanced Microstructures, Nanjing University, Nanjing 210093, China}

\author{Shengjun Wu}
\email{sjwu@nju.edu.cn}
\affiliation{National Laboratory of Solid State Microstructures, School of physics and college of engineering and applied sciences, and Collaborative Innovation Center of Advanced Microstructures, Nanjing University, Nanjing 210093, China}

\author{Lijian Zhang}
\email{lijian.zhang@nju.edu.cn}
\affiliation{National Laboratory of Solid State Microstructures, School of physics and college of engineering and applied sciences, and Collaborative Innovation Center of Advanced Microstructures, Nanjing University, Nanjing 210093, China}

\author{Penghui Yao}
\email{phyao1985@gmail.com}
\affiliation{State Key Laboratory of Novel Software Technology, Nanjing University, Nanjing 210023, China}
\affiliation{
Hefei National Laboratory, Hefei 230088, China}

\date{\today}

\begin{abstract}
Gaussian boson sampling (GBS) seeks quantum advantage by sampling photon-number patterns generated with squeezed inputs and passive linear optics. Many proposed GBS applications instead target function computation by applying functions to mode-resolved photon-number outcomes---a natural form of experimental postprocessing that produces classical outputs. Sampling hardness alone, however, does not determine the complexity of these tasks.
By analyzing the irreducible decomposition of fixed-photon-number operator spaces, we prove that the expectation value of every such function in the average case over passive linear-optical networks can be classically evaluated for inputs with finite squeezing strength. We also provide a classical algorithm that estimates this value to inverse-polynomial additive error. The result provides new theoretical tools for analyzing linear-optical quantum
systems, helps clarify the origin of current GBS hardness evidence, and inspires new applications of GBS with genuine
quantum advantages.
\end{abstract}

\maketitle

\noindent\textit{Introduction.--}
In the noisy intermediate-scale quantum (NISQ) era, quantum random sampling has emerged as a leading route to probing quantum computational advantage: such protocols can often be implemented without universal control over the quantum system~\cite{Preskill2018,AaronsonArkhipov2013,HangleiterEisert2023,Hamilton2017}. Gaussian boson
sampling (GBS) realizes this paradigm by injecting squeezed-vacuum states into
a passive linear-optical network and performing photon-number-resolving
(PNR) detection at the output~\cite{Hamilton2017,Kruse2019}. Probabilities of
photon-number patterns are governed by matrix hafnians, while threshold
detection leads to Torontonians~\cite{Hamilton2017,QuesadaThreshold2018}.
Subsequent theoretical work has strengthened the evidence for the
classical hardness of sampling within a sufficiently small
constant total-variation distance of ideal GBS output distributions
~\cite{Deshpande2022,GrierBrod2022,EhrenbergAnti2025,
Ehrenberg2025,ShouMiller2025}. In parallel, GBS experiments have advanced from early few-photon
demonstrations to systems with more than a thousand
squeezed inputs~\cite{Zhong2019,Zhong2020,ArrazolaNature2021,
Zhong2021,Madsen2022,Deng2023,LiuJiuzhang2026}.

In recent years, there have been efforts to explore the
potential quantum advantage of GBS in computational tasks in which the
sampling outcomes are used as inputs to further classical computation. In molecular spectroscopy, probabilities of photon-number patterns
can be combined to calculate Franck--Condon profiles
and thereby obtain vibronic spectra~\cite{Huh2015,HuhYung2017,
ZhuVibronic2024,Eickmann2026}. In graph applications, hafnians link these probabilities to perfect matchings, biasing samples toward dense subgraphs or large-weight cliques and motivating applications to molecular docking~\cite{Bradler2018,ArrazolaBromley2018,Sempere2022,DengGraph2023,BanchiDocking2020,YuDrug2023}. Related proposals use statistics of GBS outcomes to construct graph kernels, point processes, trainable distributions, or features for classification and image recognition~\cite{Schuld2020,Jahangiri2020,BanchiTraining2020,Anteneh2023,GongImage2026}. Although their goals differ, these applications follow the same basic
procedure: classical postprocessing takes mode-resolved PNR
outcomes as inputs to compute task-specific function values or their averages.

A similar scenario appears in standard quantum algorithms. Shor's factoring algorithm uses classical postprocessing
of its quantum circuit's measurement outcomes to infer a period and hence a factor, while Grover's algorithm uses the measured bit string to identify a marked item~\cite{Shor1997,Grover1997}. This raises the corresponding question for sampling algorithms: can they still offer a
computational advantage when the goal is function computation rather than
sampling from the full output distribution? Several works have addressed
special cases and related formulations for GBS. For molecular vibronic spectroscopy,
a quantum-inspired classical algorithm exactly solves the zero-temperature
harmonic instances corresponding to GBS~\cite{OhVibronic2024}; for graph
applications, a quantum-inspired sampler has been developed for the
densest-$k$-subgraph and maximum-weight-clique problems~\cite{OhGraph2024}. More generally, Lim and Oh give classical algorithms for
estimating expectation values in linear-optical circuits with arbitrary
product inputs and subsystem observables satisfying a polynomial
Hilbert--Schmidt norm bound~\cite{LimOh2026}; they also establish efficient classical estimation of
observable expectation values for circuits with Gaussian transformations
and a constant number of adaptive measurement-and-feedforward steps
~\cite{OhLim2026}. These results, however, leave open whether the sampling hardness of
GBS can yield a quantum advantage in function computation.
In particular, conditioned on a total photon number proportional to the
number of modes, the probability of each fixed GBS output pattern is
exponentially small on average over Haar-random
networks~\cite{Arienzo2025,Supplemental}; therefore a function-computation advantage must instead aggregate outcomes into an inverse-polynomially resolvable signal while retaining classical hardness.

In this work, we introduce an operational framework for GBS function
computation. Suppose that each GBS 
run with an $m$-mode unitary $U$ of the linear optical network produces a mode-resolved PNR outcome $\vct s=(s_1,\ldots,s_m)\in\N^m$ with probability $P_U(\vct s)$. We consider a family of real-valued outcome functions $f:\N^m\to\mathbb R$ and define
$\mu_f(U):=\E_{\vct s\sim P_U}[f(\vct s)]
=\sum_{\vct s\in\mathbb N^m}
P_U(\vct s)f(\vct s)$.
Equivalently, $\mu_f(U)=\langle\hat O_f\rangle$ is the expectation value of the 
observable $\hat O_f:=\sum_{\vct s}f(\vct s)
|\vct s\rangle\langle\vct s|$ diagonal in the photon-number basis. We consider functions which are
evaluable in polynomial time on each recorded PNR outcome~\cite{Phillips2019,Cardin2024,SeronBinned2024}. Otherwise, the hardness can simply be hidden in the classical postprocessing. To ensure that $\mu_f(U)$ can be estimated
with inverse-polynomial additive error
using polynomially many experimental samples, we require a known
bound $V_m=\poly(m)$ such that
$\operatorname{Var}_{\vct s\sim P_U}[f(\vct s)]
=\sum_{\vct s\in\mathbb N^m}
P_U(\vct s)\bigl(f(\vct s)-\mu_f(U)\bigr)^2
\le V_m$. The values of $f$ may be exponentially large while remaining
representable with a polynomial number of bits. This class of outcome functions therefore includes functions
used in many proposed GBS applications, such as vibronic energy-bin
indicators, photon-number correlators, and scores assigned to sampled
subgraphs, provided they satisfy the assumptions above~\cite{Huh2015,OhVibronic2024,Schuld2020,
BressaniniBinned2025,Phillips2019,Cardin2024,
ArrazolaBromley2018,BanchiDocking2020,BanchiTraining2020}.

We will show that the GBS function computation defined above can be evaluated classically in the average case over passive linear-optical networks for inputs with finite squeezing strength. We also construct a polynomial-time classical algorithm that estimates $\mu_f(U)$ to inverse-polynomial additive error. Together with efficient classical sampling for coherent inputs and for Gaussian states measured by homodyne detection~\cite{RahimiKeshari2016,Bartlett2002}, we identify the resource boundary summarized in Table~\ref{tab:boundary}: current GBS hardness evidence relies jointly on squeezed inputs,
mode-resolved photon detection (PNR or threshold),
and sampling from the full output distribution. Moreover, our result may inspire new GBS applications with genuine quantum advantages for practical utility.

\begin{table*}[t]
\caption{Resource boundary of standard GBS. Each of the first three rows
changes one displayed ingredient while retaining the other two; passive
interference is retained throughout. The asterisk denotes the high-probability guarantee in
Theorem~\ref{thm:main}.}
\label{tab:boundary}

\small
\renewcommand{\arraystretch}{1.12}
\setlength{\tabcolsep}{3pt}

\begin{tabular*}{\textwidth}{@{\extracolsep{\fill}}lcccc}
\toprule
Variant
& \shortstack{Squeezed\\input}
& \shortstack{Mode-resolved\\PNR or threshold}
& \shortstack{Full-distribution\\sampling}
& Known status \\
\midrule

Coherent input
& $\times$ & $\checkmark$ & $\checkmark$
& Efficient classical sampling~\cite{RahimiKeshari2016} \\

Homodyne readout
& $\checkmark$ & $\times$ & $\checkmark$
& Efficient classical sampling~\cite{Bartlett2002} \\

Function-average estimation
& $\checkmark$ & $\checkmark$ & $\times$
& \shortstack{Efficient classical estimation$^{*}$ (this work)} \\

Standard GBS
& $\checkmark$ & $\checkmark$ & $\checkmark$
& Sampling-hardness evidence~\cite{Hamilton2017,HangleiterEisert2023} \\
\bottomrule
\end{tabular*}
\end{table*}

\noindent\textit{Model and result.--}
We consider ideal GBS with lossless optics and unit-efficiency PNR detection. For $m$ modes, set
$t_j=e^{i\phi_j}\tanh r_j$ for $j=1,\ldots,m$, with
$0\le r_j\le r_{\max}<\infty$, where $r_{\max}$ is independent of $m$. We refer to this condition as finite squeezing strength. The product squeezed-vacuum
input is
\begin{equation}
\ket{\Psi_{\mathrm{in}}}
=\prod_{j=1}^{m}(1-|t_j|^2)^{1/4}
\exp\left[
-\frac12\sum_{j=1}^{m}t_j\hat a_j^{\dagger 2}
\right]\ket{0}.
\label{eq:input}
\end{equation}

Here $\hat a_j^\dagger$ is the bosonic creation operator for mode $j$, and $\ket0$ denotes the multimode vacuum. A passive network $U\in U(m)$, represented on Fock space by
$\hat{\mathcal U}(U)$, satisfies $\hat{\mathcal U}(U)\hat a_j^\dagger
\hat{\mathcal U}(U)^\dagger
=\sum_{i=1}^{m}U_{ij}\hat a_i^\dagger.$ Writing
$\ket{\Psi_U}:=\hat{\mathcal U}(U)\ket{\Psi_{\mathrm{in}}}$,
mode-resolved PNR detection produces
$\vct s=(s_1,\ldots,s_m)\in\N^m$ with probability $P_U(\vct s)=|\langle\vct s|\Psi_U\rangle|^2.$
We consider real outcome functions
$f:\mathbb N^m\to\mathbb R$ that are polynomial-time evaluable and
independent of $U$. The quantity of
interest is
$\mu_f(U):=\sum_{\vct s}f(\vct s)P_U(\vct s)$. The variance bound on $f(\vct s)$ under $P_U$, stated above,
is assumed to hold for every $U\in U(m)$. We also assume a known bound
$|\mu_f(U)|\le M_m$ for every $U$, with $\log M_m=\poly(m)$. In the Supplemental Material we relax these bounds to hold
only with high probability over Haar-random $U$~\cite{Supplemental}.

\begin{theorem}
\label{thm:main}
For the GBS function-computation model specified above with the mode number $m$ and fixed constants $c,\kappa>0$, there exists a randomized classical polynomial-time algorithm
$\mathcal A_{c,\kappa}$ that, given a Haar-random network unitary $U$, the squeezing
parameters $\{t_j\}_{j=1}^{m}$, the outcome function $f$ with the known bounds $M_m$ and $V_m$ which are independent of $U$, outputs an estimate $\widetilde{\mu}_f(U)$ satisfying
\begin{equation}
\left|\widetilde{\mu}_f(U)-\mu_f(U)\right|
\le m^{-c}
\label{eq:main}
\end{equation}
with success probability at least $1-m^{-\kappa}$.
The probability is over the Haar-random choice of $U$ and the randomness of $\mathcal A_{c,\kappa}$.
\end{theorem}

The main theorem can be proved by decomposing $\mu_f(U)$ into contributions from
sectors of fixed total photon number and then resolving each contribution
into correlation orders, a hierarchy introduced below. The Haar-random choice of the network $U$ makes the higher-order squeezed-pair correlations weakly overlap, while finite squeezing strength permits truncation at a
constant order to achieve any prescribed inverse-polynomial additive error. Retaining only the low-order contributions yields a classically computable approximation for $\mu_f(U)$. For clarity, the analysis below is presented for the normalized setting
$f:\mathbb N^m\to[-1,1]$, while the extension to the general class of outcome
function is provided in the Supplemental
Material~\cite{Supplemental}.

\noindent\textit{Operator-space decomposition and correlation-order
structure.--}
This section first introduces the mathematical setting and develops a general
framework for analyzing function-computation tasks in the average case over
passive linear-optical networks with arbitrary bosonic inputs.
The framework
is then specialized to the squeezed-vacuum inputs of GBS in the next section. 
Let
\(\mathcal H_{m,n}:=\Sym^n(\mathbb C^m)\)
be the Hilbert space of \(n\) identical photons distributed over \(m\)
modes, and let \(\hat\rho_n\) be an arbitrary normalized state on
\(\mathcal H_{m,n}\). An interferometer $U\in U(m)$ acts on this sector through
$\hat{\mathcal U}^{(n)}(U)
:=\hat{\mathcal U}(U)|_{\mathcal H_{m,n}}
=\Sym^n(U)$.
Equivalently,
$\hat{\mathcal U}(U)
=\bigoplus_{n\ge0}\hat{\mathcal U}^{(n)}(U)$. Because passive linear optics preserves total photon number and the
observables considered here are diagonal in the multimode Fock basis, the
analysis can be performed independently in each fixed photon-number sector. 
Consider the action of \(\hat{\mathcal U}^{(n)}(U)\) on the operator space $\mathcal B(\mathcal H_{m,n})$: $\hat X\mapsto
\hat{\mathcal U}^{(n)}(U)\hat X
\hat{\mathcal U}^{(n)}(U)^\dagger$,
\(\mathcal B(\mathcal H_{m,n})\) is decomposed into multiplicity-free irreducible \(U(m)\)-submodules~\cite{Arienzo2025},
\begin{equation}
 \mathcal B(\mathcal H_{m,n})
 =\bigoplus_{k=0}^{n}\mathcal V_{n,k}.
 \label{eq:irrep-decomposition}
\end{equation}
The label $k$ has a direct interpretation in terms of ordered bosonic operators. For $\vct u,\vct v\in\mathbb N^m$, write
$(\hat{\vct a}^{\dagger})^{\vct u}
:=\prod_{j=1}^{m}(\hat a_j^\dagger)^{u_j}$,
$\hat{\vct a}^{\vct v}
:=\prod_{j=1}^{m}\hat a_j^{v_j}$, and
$|\vct u|:=\sum_{j=1}^{m}u_j$. Let
$\mathcal O_{n,\le k}:=\operatorname{span}\{
(\hat{\vct a}^{\dagger})^{\vct u}
\hat{\vct a}^{\vct v}|_{\mathcal H_{m,n}}:
|\vct u|=|\vct v|\le k\}$, which is spanned by number-preserving normally ordered monomials containing
at most $k$ creation-annihilation pairs. The family $\{\mathcal O_{n,\le k}\}_{k=0}^{n}$ forms an increasing
sequence of operator spaces:
$\mathcal O_{n,\le0}\subseteq\cdots\subseteq
\mathcal O_{n,\le n}=\mathcal B(\mathcal H_{m,n})$. With respect to the Hilbert--Schmidt inner product, the component added
when the index increases from $k-1$ to $k$ is
$\mathcal V_{n,k}
=\mathcal O_{n,\le k}\cap\mathcal O_{n,\le k-1}^{\perp}
\cong\mathcal O_{n,\le k}/\mathcal O_{n,\le k-1}$. Equivalently,
$\mathcal O_{n,\le K}=\bigoplus_{k=0}^{K}\mathcal V_{n,k}$~\cite{Mhiri2026,Monbroussou2026,Supplemental}.
Thus, $k$ labels correlations first captured by normally ordered monomials
containing $k$ creation-annihilation pairs, and we refer to $k$ as the
correlation order. For example, $\mathcal V_{n,0}=\operatorname{span}\{\hat I_n\}$,
$\mathcal V_{n,1}$ is spanned by the traceless one-body operators
$\hat a_i^\dagger\hat a_j-(n/m)\delta_{ij}\hat I_n$, and
$\mathcal V_{n,2}$ consists of the two-body operators with their scalar and
one-body contractions removed. 
For $m=2$, the Schwinger-boson representation identifies
$\mathcal H_{2,n}$ with the spin-$j=n/2$ representation~\cite{Schwinger1965,Arecchi1972}. Under this identification,
$\mathcal V_{n,k}$ is precisely the rank-$k$ irreducible
spherical-tensor sector, equivalently the spin-$k$ sector, of dimension
$2k+1$~\cite{Fano1957}.

The function-computation tasks considered here are represented by observables
diagonal in the multimode Fock basis. Let $\Phi_{m,n}:=
\{\vct s\in\mathbb N^m:|\vct s|=n\}, D_{m,n}:=|\Phi_{m,n}|=\binom{m+n-1}{n},$ and let $\nu_{m,n}$ be the uniform measure on $\Phi_{m,n}$. For the function with normalized outcome $f:\mathbb N^m\to[-1,1]$, write
$f_n:=f|_{\Phi_{m,n}}$ and $\hat O_{f,n}
:=
\sum_{\vct s\in\Phi_{m,n}}
f_n(\vct s)\ket{\vct s}\!\bra{\vct s}.$
Under this identification, the $\mathcal V_{n,k}$ component of
$\hat O_{f,n}$ corresponds to the order-$k$ Hahn component $f_{n,k}$
in the orthogonal expansion of $f_n$,
\begin{equation}
f_n(\vct s)=\sum_{k=0}^{n}f_{n,k}(\vct s),\;
\hat O_{f,n,k}:=
\sum_{\vct s\in\Phi_{m,n}}
f_{n,k}(\vct s)\ket{\vct s}\!\bra{\vct s}.
\end{equation}
where $f_{n,k}$ denotes the $L^2(\nu_{m,n})$-orthogonal projection of
$f_n$ onto the order-$k$ multivariate Hahn subspace, and
$\hat O_{f,n,k}\in\mathcal V_{n,k}$
~\cite{KhareZhou2009,Xu2015}. Details of this correspondence are given in
the Supplemental Material~\cite{Supplemental}.
The function-space interpretation of the lowest correlation orders is
explicit: $k=0$ consists of the constant functions; $k=1$ is spanned by
the centered occupations $s_i-n/m$; and $k=2$ is spanned by the order-$2$ Hahn projections
of $s_i s_j$ for $i<j$. The correspondence between the Hahn components of $f_n$ and the
correlation-order components of $\hat O_{f,n}$ allows $\mu_f(U)$ to be
analyzed separately at each correlation order.

For an arbitrary normalized state $\hat\rho_n$ on
$\mathcal H_{m,n}$, let $\hat\rho_n=\sum_{k=0}^{n}\hat\rho_{n,k}, \hat\rho_{n,k}\in\mathcal V_{n,k},$ be its orthogonal decomposition and let
$\hat U_n:=\hat{\mathcal U}^{(n)}(U)$, the sector expectation is
\begin{equation}
\mu_{f,n}(U)
:=
\Tr\!\left[
\hat O_{f,n}\hat U_n\hat\rho_n\hat U_n^\dagger
\right]
=
\sum_{k=0}^{n}\mu_{f,n}^{(k)}(U),
\end{equation}
where $\mu_{f,n}^{(k)}(U)
:=
\Tr\!\left[
\hat O_{f,n,k}\hat U_n
\hat\rho_{n,k}\hat U_n^\dagger
\right].$ Its order-$K$ truncation is
$\mu_{f,n}^{(\le K)}(U):=
\sum_{k=0}^{\min\{K,n\}}\mu_{f,n}^{(k)}(U)$. Define the normalized order-$k$ input weight
$\Lambda_{\rho,n,k}:=
D_{m,n}\|\hat\rho_{n,k}\|_{\mathrm{HS}}^2/
\dim\mathcal V_{n,k}$, where
$\|X\|_{\mathrm{HS}}^2:=\Tr(X^\dagger X)$.
Since $f$ is independent of $U$, Schur orthogonality gives
$\E_U[\mu_{f,n}^{(k)}(U)\mu_{f,n}^{(\ell)}(U)]
=\delta_{k\ell}\Lambda_{\rho,n,k}
\|f_{n,k}\|_{L^2(\nu_{m,n})}^2$. Consequently,
\begin{equation}
\begin{aligned}
 &\E_{U\sim\Haar}
 \left|\mu_{f,n}(U)-\mu_{f,n}^{(\le K)}(U)\right|^2 \\
 &\qquad=
 \sum_{K<k\le n}\Lambda_{\rho,n,k}
\norm{f_{n,k}}_{L^2(\nu_{m,n})}^2
\le
\max_{K<k\le n}\Lambda_{\rho,n,k}.
\end{aligned}
\label{eq:schur-identity}
\end{equation}
Here
$\norm{f_{n,k}}_{L^2(\nu_{m,n})}^2
:=D_{m,n}^{-1}\sum_{\vct s\in\Phi_{m,n}}
|f_{n,k}(\vct s)|^2$.
The last inequality in Eq.~\eqref{eq:schur-identity}
follows from Parseval's identity and $|f(\vct s)|\le1$:
the squared norms of the components $f_{n,k}$ sum to at most one. Classical approximation by the order-$K$ truncation therefore reduces to
bounding $\max_{K<k\le n}\Lambda_{\rho,n,k}$.

\noindent\textit{Application to GBS function computation.--} The framework can now be specialized to GBS.
Since the phases can be absorbed into $U$, it suffices to take
$t_j=\tanh r_j$ real and nonnegative. Define
$\hat K^\dagger:=\frac12\sum_{j=1}^{m}t_j\hat a_j^{\dagger 2}$.
Only even photon-number sectors are occupied, so the analysis is restricted
to $n=2N$. For the $2N$-photon sector, the normalization coefficient and normalized
state are
\begin{equation}
\mathcal Z_N:=[z^N]\prod_{j=1}^m
\left(1-|t_j|^2z\right)^{-1/2},
\ket{\Omega_N}:=
\frac{(\hat K^\dagger)^N\ket{0}}{N!\sqrt{\mathcal Z_N}}.
\label{eq:sector-state}
\end{equation}
The corresponding sector weight is $q_N
:=
\prod_{j=1}^{m}(1-|t_j|^2)^{1/2}\mathcal Z_N,$
and hence $\ket{\Psi_{\mathrm{in}}}
=
\sum_{N\ge0}(-1)^N\sqrt{q_N}\ket{\Omega_N}.$ Because the observable associated with $f$ is diagonal in the multimode
Fock basis, coherences between distinct photon-number sectors do not
contribute, and
$\mu_f(U)=\sum_{N\ge0}q_N\mu_{f,2N}(U)$.
Equation~\eqref{eq:schur-identity} applies to each $2N$-photon sector with
$\hat\rho_{2N}=\ket{\Omega_N}\!\bra{\Omega_N}$.
The GBS-specific analysis therefore reduces to bounding $q_N$ and
$\max_{K<k\le2N}\Lambda_{\rho,2N,k}$, and computing
$\mu_{f,2N}^{(k)}(U)$ efficiently for
$k\le\min\{K,2N\}$.

To obtain a simple asymptotic form for the correlation-order coefficients,
consider first the equal-squeezing case $t_j=t=\tanh r$. In every fixed-$N$ sector, the normalized input state is invariant under
real orthogonal transformations of the modes,
which forces all odd correlation orders to vanish and gives
\begin{equation}
\begin{cases}
\Lambda_{\rho,2N,k}=0, & k\text{ odd},\\
\Lambda_{\rho,2N,k}\le C_k m^{-k}, & k\text{ even}.
\end{cases}
\label{eq:equal-squeezing-weights}
\end{equation}
Equation~\eqref{eq:equal-squeezing-weights} holds for each fixed correlation order $k$ and all
$N$ with $k\le 2N$; the constant $C_k$ is independent of $m$ and $N$. Related tensor-square second-moment formulas were obtained in
Ref.~\cite{Kolarovszki2026}. The exact coefficient formula and the vanishing of odd correlation orders
imply, for any fixed cutoff $K$ and $2N>K$, $\max_{K<k\le 2N}\Lambda_{\rho,2N,k}
\le C_K m^{-2(\lfloor K/2\rfloor+1)}$. By Eq.~\eqref{eq:schur-identity}, the
mean-squared truncation error over Haar-random $U$ obeys the same bound.

For general mode-dependent squeezing profiles and any fixed $B>0$,
finite squeezing allows us to retain only sectors with total photon
number $2N\le 2L_m$, where $L_m$ grows at most linearly with $m$ and
the omitted probability satisfies
$\sum_{N>L_m}q_N\le m^{-B}$. Truncating $\mu_f(U)$ simultaneously to pair
numbers $N\le L_m$ and correlation orders $k\le K$ gives
$\mu_f^{(\le K,L_m)}(U):=
\sum_{N=0}^{L_m}q_N\mu_{f,2N}^{(\le K)}(U)$.
For all squeezing profiles with $r_j\le r_{\max}$ and every fixed
$K\in\mathbb N$, the resulting error satisfies
\begin{equation}
\norm{\mu_f-\mu_f^{(\le K,L_m)}}_{L^2(U(m))}
\le C_{K,B,r_{\max}}m^{-K/2-1/4}+m^{-B}.
\label{eq:global-rms}
\end{equation}
Here $\|g\|_{L^2(U(m))}
:=(\E_{U\sim\Haar}|g(U)|^2)^{1/2}$, and
$C_{K,B,r_{\max}}$ depends only on $K$, $B$, and $r_{\max}$.
Choosing $K=\lceil 2c+\kappa+1\rceil$ and
$B=c+\kappa/2+1$, Markov's inequality applied to the squared
error shows that, for $m\ge m_0(c,\kappa,r_{\max})$,
the truncation error is at most $m^{-c}/2$ except with probability
at most $m^{-\kappa}/2$. The triangle inequality and a union bound therefore
prove Eq.~\eqref{eq:main} in the normalized setting $f:\mathbb N^m\to[-1,1]$. The extension to the
general outcome function is given in the Supplemental
Material~\cite{Supplemental}.

\noindent\textit{Classical estimation of the truncated mean.--}
The following classical algorithm efficiently estimates the truncated mean $\mu_f^{(\le K,L_m)}(U)
:=
\sum_{N=0}^{L_m}q_N
\sum_{k=0}^{\min\{K,2N\}}
\mu_{f,2N}^{(k)}(U).$ It therefore suffices to estimate each fixed-sector contribution
$\mu_{f,2N}^{(k)}(U)
=\Tr[\hat O_{f,2N,k}\hat U_{2N}\hat\rho_{2N,k}
\hat U_{2N}^{\dagger}]$ for $N\le L_m$ and
$k\le\min\{K,2N\}$. Instead of constructing $\hat\rho_{2N,k}$ explicitly on the full space
$\mathcal H_{m,2N}$, the algorithm uses the normalized $k$-particle
reduced density operator $\hat\gamma_N^{(k)}$ of $\ket{\Omega_N}$. This is
possible because the natural lifting of $\hat\gamma_N^{(k)}$ to
$\mathcal H_{m,2N}$ has a $\mathcal V_{2N,k}$ component proportional to
$\hat\rho_{2N,k}$; removing the lower-order components of $\hat\gamma_N^{(k)}$ and applying the
known normalization therefore recovers $\hat\rho_{2N,k}$~\cite{Monbroussou2026,Supplemental}. For $\vct u,\vct v\in\Phi_{m,k}$, with
$\vct u!:=\prod_{j=1}^{m}u_j!$ and
$\vct v!:=\prod_{j=1}^{m}v_j!$, the matrix elements of
$\hat\gamma_N^{(k)}$ are
\begin{equation}
\bra{\vct u}\hat\gamma_N^{(k)}
\ket{\vct v}
=
\frac{
 \bra{\Omega_N}
 (\hat{\vct a}^{\dagger})^{\vct v}
 \hat{\vct a}^{\vct u}
 \ket{\Omega_N}
}{
 \binom{2N}{k}\sqrt{\vct u!\vct v!}
}.
\label{eq:reduced-state-main}
\end{equation}
Expanding $\ket{\Omega_N}$ in the Fock basis expresses each matrix
element of $\hat\gamma_N^{(k)}$ as a sum of products of single-mode
terms. This sum can be evaluated one mode at a time while tracking the fixed total photon number $2N$. Thus all $D_{m,k}^2=\binom{m+k-1}{k}^2$ matrix elements of
$\hat\gamma_N^{(k)}$ can be computed in time polynomial in $m$
and $N$ for fixed $k$.

Since $\hat O_{f,2N,k}$ is diagonal in the multimode Fock basis,
evaluating $\mu_{f,2N}^{(k)}(U)$ requires the diagonal weight associated
with each complete $2N$-photon pattern. This weight can be reconstructed
from the evolved $k$-particle reduced state. Let
$\hat\gamma_{N,U}^{(k)}
:=\hat{\mathcal U}^{(k)}(U)\hat\gamma_N^{(k)}
\hat{\mathcal U}^{(k)}(U)^\dagger$.
For $\vct s\in\Phi_{m,2N}$, the diagonal information through correlation
order $k$ is obtained from
$\sum_{\substack{\vct u\le\vct s\\|\vct u|=k}}
\binom{\vct s}{\vct u}
\bra{\vct u}\hat\gamma_{N,U}^{(k)}\ket{\vct u}$,
where
$\binom{\vct s}{\vct u}:=\prod_{j=1}^{m}\binom{s_j}{u_j}$.
This quantity contains correlation orders $0,\ldots,k$; removing the
lower-order terms and applying the known normalization isolates
$w_{N,k}(\vct s;U)
:=\bra{\vct s}\hat U_{2N}\hat\rho_{2N,k}
\hat U_{2N}^{\dagger}\ket{\vct s}$~\cite{Supplemental}. By orthogonality between distinct correlation orders, the corresponding
order-$k$ contribution is
$\mu_{f,2N}^{(k)}(U)
=\sum_{\vct s\in\Phi_{m,2N}}
f(\vct s)w_{N,k}(\vct s;U)$~\cite{Supplemental}. Summing the retained orders gives
\begin{equation}
W_{N,K}(\vct s;U)
:=
\sum_{k=0}^{\min\{K,2N\}}w_{N,k}(\vct s;U).
\label{eq:retained-weight-main}
\end{equation}
For fixed $K$, evaluating $W_{N,K}(\vct s;U)$ uses only
$\hat\gamma_N^{(k)}$ and the matrix elements
$\bra{\vct u}\hat{\mathcal U}^{(k)}(U)\ket{\vct v}$
for $k\le\min\{K,2N\}$. These matrix elements are given by
permanents of matrices of size at most $K$, so the evaluation takes
time polynomial in $m$ and $N$.

Although $W_{N,K}(\vct s;U)$ can be evaluated efficiently for any given
pattern $\vct s$, directly summing over all
$D_{m,2N}$ patterns is generally inefficient. The outer sum over $N$ and
the inner sum over photon-number patterns are therefore evaluated
simultaneously by Monte Carlo sampling. A single sample first draws the
total pair number $N$ according to $q_N$, which can be implemented by
independently sampling the pair number in each input mode and summing the
results. If $N>L_m$, the output is zero. Otherwise, a pattern
$\vct S$ is drawn from the uniform measure $\nu_{m,2N}$ on $\Phi_{m,n}$, $f(\vct S)$ is evaluated once,
and the output is
\begin{equation}
Z_f
:=
D_{m,2N}f(\vct S)W_{N,K}(\vct S;U).
\label{eq:estimator-main}
\end{equation}
Since every pattern has probability $D_{m,2N}^{-1}$ under
$\nu_{m,2N}$, the estimator is unbiased:
$\E[Z_f]=\mu_f^{(\le K,L_m)}(U)$. Using $|f|\le1$, its second moment on every retained sector
$N\le L_m(B)$ satisfies
$\E[|Z_f|^2\mid N,U]
\le C_{K,B,r_{\max}}m^{K+1/2}$. Hence polynomially many Monte Carlo samples are sufficient.
The weight $W_{N,K}$ may nevertheless
be signed, so the algorithm estimates the prescribed outcome-function
average but neither approximates $P_U$ nor provides a classical simulation for GBS.

\noindent\textit{Discussion.--}
Table~\ref{tab:boundary} compares the present result with two established
classically simulable variants of the GBS setup: coherent inputs with passive optics yield independent Poisson counts~\cite{RahimiKeshari2016}, whereas homodyne detection of Gaussian states
yields efficiently sampleable multivariate Gaussian distributions~\cite{Bartlett2002}. The theorem in this work instead establishes a task-level resource
boundary within the ideal GBS model. Other work analyzes
physical resources such as non-Gaussianity, squeezing, and
entanglement~\cite{ChabaudWalschaers2023}, or classical simulability
under additional physical restrictions~\cite{Qi2020,
BressaniniThermal2024,QiShallow2022,GarciaPatronLossy2019,
LiuMPO2023,OhExperimental2024,LiuLowEntanglement2026}.
Faster exact samplers reduce finite-size costs of full-distribution
sampling~\cite{QuesadaArrazola2020,QuesadaSpeedup2022,Bulmer2022}.

The outcome-function class covers a broad range of nonlinear functions of the
photon-number record. For example, the number-phase function
$f(\vct s)=\cos(\pi\sum_{i=1}^{m}s_i^2/p_i)$, with polynomially specified
primes $p_i$, is bounded, independent of $U$, and efficiently evaluable; its
average is therefore covered by Theorem~\ref{thm:main}~\cite{Monbroussou2026}. This does not resolve the fixed-Fock-input
worst-case setting considered in Ref.~\cite{Monbroussou2026}, for which no
hardness result is known. The theorem's broad function coverage assumes the standard GBS setup;
proposals with altered input preparation or optimization over
repeated samples require separate analysis~\cite{Borghi2025,
Anteneh2023,ArrazolaOptimization2018,Bradler2018}.

The Haar-average setting and the independence of $f$ from $U$ are essential to the proof.
If $f$ depends on $U$, its correlation with the network
invalidates the Schur orthogonality argument in
Eq.~\eqref{eq:schur-identity}; similarly, Haar concentration provides no
guarantee for worst-case networks.  Arbitrary
$U$-dependent postprocessing and structured worst-case networks therefore
remain open and provide natural candidates for seeking quantum advantage
in GBS function computation. It is natural to ask whether analogous classical-estimation results hold
for sampling algorithms with different inputs from GBS. In separate work, we establish such results for single-photon inputs in boson sampling, arbitrary polynomial-energy Fock-state inputs, and core states with polynomial Fock support.

\begin{acknowledgments}
This work is supported by the National Natural Science Foundation
of China (Grant Nos. 12475020, 92565111, U24A2017, 12347104, 12461160276 and 62332009),
the NSFC/RGC Joint Research Scheme
(Grant No. 12461160276), 
the Quantum Science and Technology---National
Science and Technology Major Project (Grant Nos. 2021ZD0301701,
2024ZD0300900, and 2021ZD0302901),
the National Key Research and Development Program of China (Grant No. 2023YFC2205802),
the Natural Science Foundation of Jiangsu Province
(Grant Nos. BK20243060 and BK20233001), 
the Fundamental and Interdisciplinary Disciplines Breakthrough Plan of the Ministry of Education of China (Grant Nos. JYB2025XDXM105 and
JYB2025XDXM118), 
the ``111 Center'' (No. B26023), 
and the Fundamental Research Funds for the Central Universities
(Grant No. 2026300376).
\end{acknowledgments}

\nocite{Howe1989,Scheel2004,BrentZimmermann2010}
\bibliography{references}

\end{document}

\title{Supplemental Material for ``Classical Algorithms for Function Computation in Gaussian Boson Sampling''}

\author{Ruoting Dou}
\affiliation{National Laboratory of Solid State Microstructures, School of physics and college of engineering and applied sciences, and Collaborative Innovation Center of Advanced Microstructures, Nanjing University, Nanjing 210093, China}

\author{Hao Zhan}
\affiliation{National Laboratory of Solid State Microstructures, School of physics and college of engineering and applied sciences, and Collaborative Innovation Center of Advanced Microstructures, Nanjing University, Nanjing 210093, China}

\author{Shengjun Wu}
\email{sjwu@nju.edu.cn}
\affiliation{National Laboratory of Solid State Microstructures, School of physics and college of engineering and applied sciences, and Collaborative Innovation Center of Advanced Microstructures, Nanjing University, Nanjing 210093, China}

\author{Lijian Zhang}
\email{lijian.zhang@nju.edu.cn}
\affiliation{National Laboratory of Solid State Microstructures, School of physics and college of engineering and applied sciences, and Collaborative Innovation Center of Advanced Microstructures, Nanjing University, Nanjing 210093, China}

\author{Penghui Yao}
\email{phyao1985@gmail.com}
\affiliation{State Key Laboratory of Novel Software Technology, Nanjing University, Nanjing 210023, China}
\affiliation{
Hefei National Laboratory, Hefei 230088, China}
\date{\today}
\maketitle

This Supplemental Material proves the average-case classical-estimation
theorem stated in the main text.  We first establish the result for an
outcome function $f_m:\N^m\to[-1,1]$: the fixed-photon-number operator space
is decomposed according to correlation orders, the high-order components of a
squeezed-vacuum input are bounded uniformly, and the retained components are
estimated by a classical Monte Carlo algorithm.  We then extend this bounded
result to the general outcome-function class of the main text by estimating a
polynomial-size family of bounded characteristic functions and unwrapping
their phases.  The construction estimates one prescribed expectation value.
Its retained weight is generally signed and does not define an approximate
GBS output distribution.

The fixed-particle multiplicity-free decomposition and its dimensions are
imported from Ref.~\cite{Arienzo2025}; the deletion--reinsertion realization,
recursive projectors, and Schur second-moment identity are imported from
Ref.~\cite{Mhiri2026}.  The spectrum of the P\'olya down--up chain and the
multivariate Hahn spaces are standard results from
Refs.~\cite{KhareZhou2009,Xu2015}.  The general low-order surrogate principle
was developed in Ref.~\cite{Monbroussou2026}.  We restate these ingredients in
the normalization used here and prove the required bridge between the full
operator decomposition and its PNR-diagonal restriction.  The profile-uniform
correlation bound, its optimized reduced-purity envelope, the nonuniform
fixed-order estimator, and the characteristic-function extension are proved
below.

\section{Scope, conventions, and precise statement}
\label{sec:scope}

The number of optical modes is $m$. Since the squeezed-vacuum GBS input
has support only on even-photon-number sectors, we denote its photon-pair
number by $N$ and the corresponding total photon number by $n=2N$.
A deterministic PNR record is $\vct s=(s_1,\ldots,s_m)$, while $\vct S$
denotes the corresponding random record. The index $k$ is the correlation
order in a fixed-$n$ sector, and $K$ is a correlation-order cutoff
independent of $m$. The pair-number cutoff is denoted by $L_m$.
When its dependence on the physical-tail exponent matters, we write
$L_m=L_m(B)$ with $B>0$.
We use the rising and falling factorials
$\Poch{a}{r}:=\prod_{j=0}^{r-1}(a+j)$ and
$\fall{a}{r}:=\prod_{j=0}^{r-1}(a-j)$, respectively, with both empty
products equal to $1$ when $r=0$.

For a family $f_m:\N^m\to\R$, \emph{efficiently evaluable} means that a
classical algorithm, given the binary encoding of $\vct s$ and a precision
parameter $p$, returns a rational approximation to $f_m(\vct s)$ with
absolute error at most $2^{-p}$ in time polynomial in $m$, the encoding
length of $\vct s$, and $p$.  The returned rational has encoding length
polynomial in the same parameters; this is the bit-model realization of the
polynomial-bit-output assumption in the main text.  The
program defining the family is part of the specification of the task.  As in
the main text, $f_m$ is fixed independently of the Haar draw of the
interferometer, although it may depend on $m$ and on problem data fixed before
that draw.
The main text suppresses the family subscript and writes $f$ for $f_m$.

For the GBS output law $P_U$, write
\begin{equation}
 \mu_f(U):=\E_{\vct S\sim P_U}[f_m(\vct S)].
 \label{eq:mu-f-definition-supp}
\end{equation}

This outcome-function formulation is intended to capture the familiar
measurement-and-decoding paradigm of standard quantum algorithms, including
Shor's factoring algorithm and Grover search.  More precisely, let
$\mathcal D_m:\N^m\to\mathcal Y_m$ be a polynomial-time classical decoder
fixed independently of $U$, and let
$h_m:\mathcal Y_m\to\R$ be an efficiently evaluable score.  The choice
\begin{equation}
 f_m(\vct s)
 :=
 h_m\!\left(\mathcal D_m(\vct s)\right)
 \label{eq:decoder-composition}
\end{equation}
gives
\begin{equation}
 \mu_f(U)
 =
 \E_{\vct S\sim P_U}
 \!\left[
 h_m\!\left(\mathcal D_m(\vct S)\right)
 \right].
 \label{eq:decoder-expectation}
\end{equation}
Thus prescribed statistics of an efficiently decoded classical output,
including acceptance and success probabilities, are outcome-function
averages of the form considered here.  In Shor's algorithm, measurement
outcomes are processed using continued fractions and
greatest-common-divisor computations to infer a period and hence a factor,
whereas in Grover search the measured bit string is interpreted and checked
as a candidate marked item~\cite{Shor1997,Grover1997}.  We use these examples
only to identify the common measurement-plus-classical-decoding paradigm and do not claim that the Shor or Grover circuits are instances of GBS, or
that the classical algorithm proved here simulates either algorithm.

The phrase ``in the average case over passive linear-optical networks'' is
used with the following precise quantifier.  For every fixed $\lambda>0$,
there are known bounds $M_m=M_m(\lambda)\ge1$ and
$V_m=V_m(\lambda)\ge0$ satisfying
\begin{equation}
 \log M_m=\poly(m),\qquad V_m=\poly(m),
 \label{eq:mean-variance-scales}
\end{equation}
such that the Haar-good set
\begin{equation}
 \mathcal G_m(M_m,V_m)
 :=\left\{U:\ |\mu_f(U)|\le M_m,\ 
 \operatorname{Var}_{\vct S\sim P_U}[f_m(\vct S)]\le V_m\right\}
 \label{eq:Haar-good-set}
\end{equation}
obeys
\begin{equation}
 \Prob_{U\sim\Haar}[U\notin\mathcal G_m(M_m,V_m)]\le m^{-\lambda}.
 \label{eq:Haar-good-set-probability}
\end{equation}
The bounds $M_m(\lambda)$ and $V_m(\lambda)$ are either supplied as part of
the task specification with polynomial-bit encodings, or can be computed in
polynomial time from $m$, $\lambda$, and the fixed task specification.  In
either case their encoded values are available to the algorithm.
The dependence on $\lambda$ is suppressed below.  This formulation also
covers the case in which the two bounds hold for every $U$.  Haar-averaged
moment assumptions can imply this good-set formulation by Markov's
inequality, but the required moments of the mean and of the conditional
variance must then be stated separately.

We use the standard bit-oracle input model for the optical data.  On request
for $p$ bits, the real and imaginary parts of the entries of $U$ and of the
squeezing amplitudes are returned within $2^{-p}$ in time polynomial in
$m$ and $p$; the exact matrix represented by the oracle is promised to be
unitary.  Section~\ref{sec:finite-precision} shows that only polynomially many
bits are required.

\begin{theorem}[Average-case classical estimation]
\label{thm:main-supp}
Let $t_j=e^{i\phi_j}\tanh r_j$ and assume
\begin{equation}
 0\le r_j\le r_{\max}<\infty
 \quad\text{for all }j,m,
 \label{eq:bounded-squeezing}
\end{equation}
where $r_{\max}$ is independent of $m$.  Let
$f_m:\N^m\to\R$ be an efficiently evaluable family, independent of $U$,
with polynomial-bit outputs, and suppose the average-case bounds
Eqs.~\eqref{eq:mean-variance-scales}--\eqref{eq:Haar-good-set-probability}
are known.
For every fixed $c,\kappa>0$, there is a randomized classical
polynomial-time algorithm $\mathcal A_{c,\kappa}$ which, given $m$, $U$, the
squeezing parameters $t_1,\ldots,t_m$, the evaluation procedure for $f_m$,
and access to the bounds
$M_m(\kappa+2),V_m(\kappa+2)$ as specified above, outputs
$\widetilde\mu_f(U)$ such that, for all sufficiently large $m$,
\begin{equation}
 \Prob_{U\sim\Haar,\,\mathcal A}
 \left[
 \abs{\widetilde\mu_f(U)-\mu_f(U)}>m^{-c}
 \right]
 \le m^{-\kappa}.
 \label{eq:main-supp-guarantee}
\end{equation}
The running-time exponent may depend on
$c,\kappa,r_{\max}$, the polynomial bounds in
Eq.~\eqref{eq:mean-variance-scales}, and the evaluation cost of $f_m$, but it is
uniform over all squeezing profiles satisfying Eq.~\eqref{eq:bounded-squeezing},
including zero entries, degeneracies, and arbitrary phases.
\end{theorem}

Sections~\ref{sec:photon-sectors}--\ref{sec:classical-estimator} first prove
the result under the additional normalization
$f_m:\N^m\to[-1,1]$.  Section~\ref{sec:general-functions} then removes this
normalization without introducing a second generic outcome-function symbol.
Physical states and PNR-diagonal observables carry hats throughout; functions
and their expectation values do not.

\section{Squeezed input and photon-number sectors}
\label{sec:photon-sectors}

\subsection{Fock conventions and exact sector law}

Let
\begin{equation}
 \mathcal H_{m,n}:=\Sym^n(\C^m),
 \qquad
 D_{m,n}:=\dim\mathcal H_{m,n}=\binom{m+n-1}{n}.
 \label{eq:Dmn}
\end{equation}
We set $D_{m,-1}:=0$.  The representation-theoretic formulas below are
stated for $m\ge2$; for $m=1$, every fixed-photon-number space is
one-dimensional and the theorem is immediate.
For $\abs{\vct s}=n$ set
\begin{equation}
 \ket{\vct s}
 :=\prod_{j=1}^m
 \frac{(\hat a_j^\dagger)^{s_j}}{\sqrt{s_j!}}\ket0,
 \qquad
 \vct s!:=\prod_{j=1}^m s_j!.
 \label{eq:fock-basis}
\end{equation}
A passive interferometer is represented on Fock space by
$\hat{\mathcal U}(U)$ and on $\mathcal H_{m,n}$ by
$\hat U_n:=\hat{\mathcal U}^{(n)}(U)=\Sym^n(U)$.  For compactness in the
technical derivations, we write its conjugation action as
\begin{equation}
 \mathscr C_n(U)[X]
 :=\hat U_nX\hat U_n^\dagger.
 \label{eq:conjugation-action}
\end{equation}

Set $x_j:=|t_j|^2$ and
\begin{equation}
 \hat K^\dagger
 :=\frac12\sum_{j=1}^m t_j\hat a_j^{\dagger2},
 \qquad
 \mathcal Z_N(\vct x)
 :=[z^N]\prod_{j=1}^m(1-x_jz)^{-1/2}.
 \label{eq:Kt-ZN}
\end{equation}
Whenever $\mathcal Z_N(\vct x)>0$, we use the main-text notation
\begin{equation}
 \mathcal Z_N:=\mathcal Z_N(\vct x),\qquad
 \ket{\Omega_N}
 :=\frac{(\hat K^\dagger)^N\ket0}
 {N!\sqrt{\mathcal Z_N}},
 \qquad
 \hat\rho_{2N}
 :=\ket{\Omega_N}\!\bra{\Omega_N}.
 \label{eq:sector-state-supp}
\end{equation}
The sign $(-1)^N$ inherited from the squeezed-vacuum exponential is an
irrelevant global phase within the $N$-pair sector.  Direct expansion gives
the pair-number law
\begin{equation}
 q_N
 =\left(\prod_{j=1}^m\sqrt{1-x_j}\right)
  \mathcal Z_N,
 \label{eq:pair-law}
\end{equation}
whose generating function is
\begin{equation}
 \sum_{N\ge0}q_Nz^N
 =\prod_{j=1}^m
 \left(\frac{1-x_j}{1-zx_j}\right)^{1/2}.
 \label{eq:pair-pgf}
\end{equation}
Thus $N=\sum_jN_j$ in distribution, where the $N_j$ are independent and
\begin{equation}
 \Prob[N_j=q]
 =\sqrt{1-x_j}\,
 \frac{\Poch{1/2}{q}}{q!}x_j^q,
 \qquad q\in\N.
 \label{eq:single-mode-NB}
\end{equation}
If $q_N=0$, the corresponding sector is never sampled and its conditional
state need not be defined.  In every global sum below we adopt the convention
$q_N\mu_{f,2N}(U)=q_N\mu_{f,2N}^{(\le K)}(U)=0$ for such a sector.
In particular,
\begin{equation}
 \E N=\frac12\sum_{j=1}^m\frac{x_j}{1-x_j}
 =\frac12\sum_{j=1}^m\sinh^2r_j=O(m).
 \label{eq:mean-pair-number}
\end{equation}
Because passive optics and PNR-diagonal observables preserve total photon
number, every output expectation is the classical mixture of the conditional
expectations in the occupied states
$\mathscr C_{2N}(U)[\hat\rho_{2N}]$.

\subsection{Physical cutoff and a uniform saddle gap}

Let $x_\star=\tanh^2r_{\max}$.  The case $x_\star=0$ is the all-vacuum
input and is immediate, so assume $0<x_\star<1$.  Choose constants
\begin{equation}
 1<z_1<z_0<x_\star^{-1},
 \qquad
 \delta_{\rm fg}:=1-z_0x_\star>0,
 \qquad
 c_0:=\frac{z_0-z_1}{z_0}>0,
 \label{eq:z-constants}
\end{equation}
and define
\begin{equation}
 x_{\max}:=\max_jx_j,
 \qquad
 \mu(s):=\frac12\sum_{j=1}^m\frac{s x_j}{1-sx_j}.
 \label{eq:tilted-mean}
\end{equation}

\begin{lemma}[Physical pair-number cutoff]
\label{lem:physical-cutoff}
For every fixed $B>0$, let
\begin{equation}
 L_m(B):=\left\lceil
 \mu(z_1)+\frac{B}{\log z_1}\log m
 \right\rceil.
 \label{eq:physical-cutoff}
\end{equation}
Then $L_m(B)=O(m)$ and
\begin{equation}
 \Prob[N>L_m(B)]\le m^{-B}.
 \label{eq:pair-tail}
\end{equation}
Moreover, for all sufficiently large $m$, either
$L_m(B)=O_B(\log m)$, or every occupied sector
$1\le N\le L_m(B)$ has a unique saddle $s_N$ satisfying
\begin{equation}
 \mu(s_N)=N,
 \qquad
 s_Nx_{\max}\le1-\delta_{\rm fg}.
 \label{eq:fugacity-gap}
\end{equation}
\end{lemma}

\begin{proof}
From Eq.~\eqref{eq:pair-pgf},
\begin{equation}
 \log\E[z_1^N]
 =\int_1^{z_1}\frac{\mu(t)}{t}\dd t
 \le\mu(z_1)\log z_1.
 \label{eq:mgf-bound}
\end{equation}
Markov's inequality applied to $z_1^N$ gives
$\Prob[N>L]\le z_1^{\mu(z_1)-L}$, proving Eq.~\eqref{eq:pair-tail}.
The bound $L_m=O(m)$ follows from
\begin{equation}
 \frac{2\mu(z_1)}m
 \le\frac{z_1x_\star}{1-z_1x_\star}.
 \label{eq:finite-density}
\end{equation}
For the dichotomy, put
$H_m:=(B/\log z_1)\log m+1$.  Direct subtraction gives
\begin{align}
 \mu(z_0)-\mu(z_1)
 &=\frac12\sum_j
 \frac{(z_0-z_1)x_j}
 {(1-z_0x_j)(1-z_1x_j)}\notag\\
 &\ge c_0\mu(z_0).
 \label{eq:mu-gap}
\end{align}
If $\mu(z_0)\le H_m/c_0$, then
$\mu(z_1)\le\mu(z_0)$ and Eq.~\eqref{eq:physical-cutoff} gives
$L_m=O_B(\log m)$.  Otherwise Eq.~\eqref{eq:mu-gap} implies
$\mu(z_0)-\mu(z_1)>H_m$, and therefore
$L_m\le\mu(z_1)+H_m<\mu(z_0)$.  Monotonicity of
$\mu$ then implies $s_N\le z_0$ for every occupied $N\le L_m$, which proves
Eq.~\eqref{eq:fugacity-gap}.  Existence and uniqueness of $s_N$ below the
first pole follow from strict monotonicity on every occupied nonvacuum
profile.
\end{proof}

\section{Correlation orders in a fixed photon-number sector}
\label{sec:correlation-orders}

\subsection{Deletion, reinsertion, and irreducible modules}

For $\vct u,\vct v\in\N^m$, write
\begin{equation}
 (\hat{\vct a}^{\dagger})^{\vct u}
 :=\prod_{j=1}^m(\hat a_j^\dagger)^{u_j},
 \qquad
 \hat{\vct a}^{\vct v}:=\prod_{j=1}^m\hat a_j^{v_j},
 \qquad
 |\vct u|:=\sum_{j=1}^m u_j,
 \label{eq:multi-index-monomials}
\end{equation}
and define the number-preserving monomial space
\begin{equation}
 \mathcal O_{n,\le k}
 :=\operatorname{span}\!\left\{
 (\hat{\vct a}^{\dagger})^{\vct u}
 \hat{\vct a}^{\vct v}\big|_{\mathcal H_{m,n}}:
 |\vct u|=|\vct v|\le k
 \right\}.
 \label{eq:monomial-spaces}
\end{equation}
We adopt the convention $\mathcal O_{n,\le-1}:=\{0\}$.
These spaces form the increasing sequence used in the main text.

We use the Hilbert--Schmidt inner product and norm
\begin{equation}
 \langle X,Y\rangle_{\HS}:=\Tr(X^\dagger Y),
 \qquad
 \norm{X}_{\HS}^2:=\langle X,X\rangle_{\HS}.
 \label{eq:HS-definition}
\end{equation}

For $n\ge1$, define the $U(m)$-equivariant maps
\begin{equation}
 \mathsf L_n(X):=\sum_{j=1}^m\hat a_jX\hat a_j^\dagger,
 \qquad
 \mathsf R_{n-1}(Y):=\sum_{j=1}^m\hat a_j^\dagger Y\hat a_j.
 \label{eq:LR-maps}
\end{equation}
They are Hilbert--Schmidt adjoints and, for $q\ge1$, obey
\begin{equation}
 \mathsf L_{q+1}\mathsf R_q-
 \mathsf R_{q-1}\mathsf L_q
 =(m+2q)\operatorname{Id}_{\rm op}.
 \label{eq:LR-commutator}
\end{equation}
Set
\begin{equation}
 \mathcal P_0:=\mathcal B(\mathcal H_{m,0}),
 \qquad
 \mathcal P_k:=\ker\mathsf L_k\quad(k\ge1),
 \qquad
 \mathcal V_{n,k}:=\mathsf R_{k\to n}(\mathcal P_k),
 \label{eq:primitive-spaces}
\end{equation}
where
$\mathsf R_{k\to n}:=\mathsf R_{n-1}\circ\cdots\circ\mathsf R_k$ and
$\mathsf L_{n\to k}:=\mathsf L_{k+1}\circ\cdots\circ\mathsf L_n$.
When $k=n$, both empty compositions are the identity map.
The known fixed-particle decomposition from
Refs.~\cite{Arienzo2025,Mhiri2026} is
\begin{equation}
 \mathcal B(\mathcal H_{m,n})
 =\bigoplus_{k=0}^n\mathcal V_{n,k},
 \label{eq:operator-decomposition}
\end{equation}
orthogonally in Hilbert--Schmidt geometry.  The modules are irreducible and
pairwise inequivalent under the conjugation action in
Eq.~\eqref{eq:conjugation-action}, with dimensions
\begin{equation}
 d_{m,k}
 :=\dim\mathcal V_{n,k}
 =D_{m,k}^2-D_{m,k-1}^2
 =\frac{m+2k-1}{m-1}\binom{m+k-2}{k}^2.
 \label{eq:module-dimension}
\end{equation}
The dimension is independent of the ambient particle number $n\ge k$.
Moreover,
\begin{equation}
 \mathcal O_{n,\le k}=\bigoplus_{j=0}^{k}\mathcal V_{n,j},
 \label{eq:monomial-irrep-correspondence}
\end{equation}
so $\mathcal V_{n,k}$ is precisely the component first captured by normally
ordered monomials containing $k$ creation--annihilation pairs.
Let $\mathcal Q_{n,k}$ denote the orthogonal projector onto
$\mathcal V_{n,k}$.

For later use, define
\begin{equation}
 \mathsf D_{n,\ell}:=
 \mathsf R_{\ell\to n}\mathsf L_{n\to\ell}.
 \label{eq:Dnell}
\end{equation}
The ladder identities in Ref.~\cite{Mhiri2026} give
\begin{equation}
 \mathsf D_{n,\ell}\big|_{\mathcal V_{n,k}}
 =\alpha_{n,\ell;k}\operatorname{Id}_{\rm op},
 \qquad
 \alpha_{n,\ell;k}
 =\frac{(n-k)!}{(\ell-k)!}
  \Poch{m+\ell+k}{n-\ell}
 \label{eq:alpha-eigenvalue}
\end{equation}
for $k\le\ell$, and zero for $k>\ell$.  In particular, the projectors
$\mathcal Q_{n,k}$ can be reconstructed triangularly from the maps
$\mathsf D_{n,\ell}$.

\subsection{Fock-basis diagonal operators and Hahn components}

Let
\begin{equation}
 \Phi_{m,n}:=\{\vct s\in\N^m:\abs{\vct s}=n\},
 \qquad
 \nu_{m,n}(\vct s):=D_{m,n}^{-1}.
 \label{eq:composition-law}
\end{equation}
For a function $f_n:\Phi_{m,n}\to\C$, define its diagonal operator
\begin{equation}
 \hat O_{f,n}:=\sum_{\vct s\in\Phi_{m,n}}
 f_n(\vct s)\ket{\vct s}\!\bra{\vct s}.
 \label{eq:diagonal-embedding}
\end{equation}
Then
\begin{equation}
 \norm{\hat O_{f,n}}_{\HS}^2
 =D_{m,n}\norm{f_n}_{L^2(\nu_{m,n})}^2.
 \label{eq:diagonal-isometry}
\end{equation}
Here
\begin{equation}
 \norm{f_n}_{L^2(\nu_{m,n})}^2
 :=\frac1{D_{m,n}}
 \sum_{\vct s\in\Phi_{m,n}}|f_n(\vct s)|^2.
 \label{eq:function-L2-norm}
\end{equation}

For $n\ge1$, consider the one-ball down--up chain on $\Phi_{m,n}$: from $\vct s$, delete
one photon from mode $j$ with probability $s_j/n$, then insert it into mode
$i$ with probability $(s_i-\delta_{ij}+1)/(m+n-1)$.  Its Markov operator is
\begin{equation}
 (\mathsf T_{m,n}f)(\vct s)
 =\sum_{i,j=1}^m
 \frac{s_j}{n}
 \frac{s_i-\delta_{ij}+1}{m+n-1}
 f(\vct s-\vct e_j+\vct e_i).
 \label{eq:Polya-chain}
\end{equation}
It is reversible with respect to $\nu_{m,n}$.  Its order-$k$ eigenspace is
the order-$k$ multivariate Hahn space, and its eigenvalue is
\begin{equation}
 \theta_{m,n,k}
 =1-\frac{k(m+k-1)}{n(m+n-1)},
 \qquad 0\le k\le n,
 \label{eq:Hahn-eigenvalue}
\end{equation}
with multiplicity
$\eta_{m,k}=\binom{m+k-2}{k}$
\cite{KhareZhou2009,Xu2015}.  Let $\mathsf H_{n,k}$ be the orthogonal
projector onto that order-$k$ space, and write
$f_{n,k}:=\mathsf H_{n,k}f_n$ and
$\hat O_{f,n,k}:=\sum_{\vct s}f_{n,k}(\vct s)
\ket{\vct s}\!\bra{\vct s}$.
For $n=0$, set $\mathsf H_{0,0}:=\operatorname{Id}$; the vacuum sector has
only the order-zero component.

\begin{proposition}[PNR-diagonal restriction]
\label{prop:Hahn-bridge}
For every $f_n:\Phi_{m,n}\to\C$ and $0\le k\le n$,
\begin{equation}
 \mathcal Q_{n,k}(\hat O_{f,n})=\hat O_{f,n,k}.
 \label{eq:Hahn-bridge}
\end{equation}
Consequently, the sum of correlation orders $k\le K$ on the diagonal PNR
subspace is the sum of the first $K+1$ Hahn components.  More explicitly,
its function-space image is
\begin{equation}
 \operatorname{span}\!\left\{
 \prod_{i=1}^m\fall{s_i}{r_i}:\
 \vct r\in\N^m,\ |\vct r|\le K
 \right\}.
 \label{eq:factorial-monomial-span}
\end{equation}
\end{proposition}

\begin{proof}
The claim for $n=0$ is immediate, so assume $n\ge1$.
The diagonal subspace is the fixed-point space of conjugation by the mode
torus.  Since every $\mathcal Q_{n,k}$ commutes with passive conjugation, it
preserves that subspace.  It remains to identify its restriction.  Applying
Eq.~\eqref{eq:LR-maps} to a diagonal matrix unit gives
\begin{align}
 \mathsf L_n(\ket{\vct s}\!\bra{\vct s})
 &=\sum_j s_j
 \ket{\vct s-\vct e_j}\!\bra{\vct s-\vct e_j},\notag\\
 \mathsf R_{n-1}(\ket{\vct y}\!\bra{\vct y})
 &=\sum_i(y_i+1)
 \ket{\vct y+\vct e_i}\!\bra{\vct y+\vct e_i}.
 \label{eq:diagonal-LR}
\end{align}
Therefore
\begin{equation}
 \mathsf R_{n-1}\mathsf L_n(\hat O_{f,n})
 =n(m+n-1)\hat O_{\mathsf T_{m,n}f,n}.
 \label{eq:LR-Polya}
\end{equation}
The eigenvalues in Eq.~\eqref{eq:Hahn-eigenvalue} are distinct.  On the full
operator space, Eq.~\eqref{eq:alpha-eigenvalue} with $\ell=n-1$ gives the
same distinct eigenvalue labels, up to the common scalar
$n(m+n-1)$.  Hence the spectral projector of
$\mathsf R_{n-1}\mathsf L_n$ onto $\mathcal V_{n,k}$ restricts to the
spectral projector of $\mathsf T_{m,n}$ onto the order-$k$ Hahn space, proving
Eq.~\eqref{eq:Hahn-bridge}.  Finally, averaging a normally ordered monomial
over the mode torus removes it unless its creation and annihilation
multi-indices agree.  For the surviving monomial,
\begin{equation}
 \bra{\vct s}
 (\hat{\vct a}^\dagger)^{\vct r}\hat{\vct a}^{\vct r}
 \ket{\vct s}
 =\prod_i\fall{s_i}{r_i}.
\end{equation}
Together with Eq.~\eqref{eq:monomial-irrep-correspondence}, this proves
Eq.~\eqref{eq:factorial-monomial-span}.
\end{proof}

\subsection{Exact Haar covariance and the state coefficients}

For a normalized state $\hat\rho_n$ on $\mathcal H_{m,n}$, write
$\hat\rho_{n,k}:=\mathcal Q_{n,k}(\hat\rho_n)$ and define, using the
notation of the main text,
\begin{equation}
 \Lambda_{\rho,n,k}
 :=D_{m,n}\norm{\hat\rho_{n,k}}_{\HS}^2\big/d_{m,k}.
 \label{eq:Lambda-definition}
\end{equation}
This dimension-normalized Hilbert--Schmidt weight measures the strength of
the order-$k$ component of the input state on its irreducible operator
subspace.
Let $f_n:=f_m|_{\Phi_{m,n}}$ and
$f_{n,k}:=\mathsf H_{n,k}f_n$.  Its corresponding operator is
\begin{equation}
 \hat O_{f,n,k}
 :=\sum_{\vct s\in\Phi_{m,n}}
 f_{n,k}(\vct s)
 \ket{\vct s}\!\bra{\vct s}.
 \label{eq:sector-readout-component}
\end{equation}
The order-$k$ contribution, total sector mean, and truncation are
\begin{align}
 \mu_{f,n}^{(k)}(U)
 &:=\Tr\!\left[
 \hat O_{f,n,k}
 \mathscr C_n(U)[\hat\rho_{n,k}]
 \right],\notag\\
 \mu_{f,n}(U)
 &:=\Tr\!\left[
 \hat O_{f,n}\mathscr C_n(U)[\hat\rho_n]
 \right]
 =\sum_{k=0}^{n}\mu_{f,n}^{(k)}(U),\notag\\
 \mu_{f,n}^{(\le K)}(U)
 &:=\sum_{k=0}^{\min\{K,n\}}\mu_{f,n}^{(k)}(U).
 \label{eq:sector-contributions}
\end{align}

\begin{theorem}[Exact Haar second moment]
\label{thm:Schur-identity}
For every $n\ge0$, every normalized state $\hat\rho_n$ on
$\mathcal H_{m,n}$, and every
$f_m:\N^m\to[-1,1]$ fixed independently of the Haar draw,
\begin{equation}
 \E_{U\sim\Haar}
 \abs{\mu_{f,n}(U)-\mu_{f,n}^{(\le K)}(U)}^2
 =\sum_{K<k\le n}\Lambda_{\rho,n,k}
 \norm{f_{n,k}}_{L^2(\nu_{m,n})}^2.
 \label{eq:exact-Schur-identity}
\end{equation}
In particular,
\begin{equation}
 \E_U
 \abs{\mu_{f,n}-\mu_{f,n}^{(\le K)}}^2
 \le\max_{K<k\le n}\Lambda_{\rho,n,k}.
 \label{eq:uniform-readout-reduction}
\end{equation}
Here and throughout, an empty maximum is defined to be zero.
\end{theorem}

\begin{proof}
For $X_k,Y_k\in\mathcal V_{n,k}$ and
$X'_{\ell},Y'_{\ell}\in\mathcal V_{n,\ell}$, Schur matrix-coefficient
orthogonality gives
\begin{align}
 &\int_{U(m)}
 \langle X_k,\mathscr C_n(U)[Y_k]\rangle_{\HS}
 \overline{\langle X'_{\ell},
 \mathscr C_n(U)[Y'_{\ell}]\rangle_{\HS}}\dd U\notag\\
 &\quad=\delta_{k\ell}
 \frac{\langle X_k,X'_{\ell}\rangle_{\HS}
       \langle Y'_{\ell},Y_k\rangle_{\HS}}
      {d_{m,k}},
 \label{eq:Schur-four-vector}
\end{align}
the specialization used in Ref.~\cite{Mhiri2026}.  Expand the truncation
error into its pairwise inequivalent components and apply
Eq.~\eqref{eq:Schur-four-vector}.  Proposition~\ref{prop:Hahn-bridge} and
Eq.~\eqref{eq:diagonal-isometry} give
\begin{equation}
 \norm{\hat O_{f,n,k}}_{\HS}^2
 =D_{m,n}\norm{f_{n,k}}_{L^2(\nu_{m,n})}^2,
 \end{equation}
which proves Eq.~\eqref{eq:exact-Schur-identity}.  Parseval and
$\norm{f_n}_\infty\le1$ imply
\begin{equation}
 \sum_{k=0}^{n}
 \norm{f_{n,k}}_{L^2(\nu_{m,n})}^2
 =\norm{f_n}_{L^2(\nu_{m,n})}^2\le1,
\end{equation}
proving
Eq.~\eqref{eq:uniform-readout-reduction}.
\end{proof}

This is the precise source--readout separation used throughout the proof:
the input fixes $\Lambda_{\rho,n,k}$, while the PNR readout contributes only
its normalized Hahn-component norm.

\subsection{Scale of a PNR pattern fixed independently of the network}

\begin{proposition}[Haar scale of one output pattern]
\label{prop:one-pattern}
For any state $\hat\rho$ on $\mathcal H_{m,n}$ and every pattern
$\vct s\in\Phi_{m,n}$ fixed independently of $U$,
define
$P_{U,\hat\rho}^{(n)}(\vct s)
:=\bra{\vct s}\mathscr C_n(U)[\hat\rho]\ket{\vct s}$.
\begin{equation}
 \E_U\bra{\vct s}\mathscr C_n(U)[\hat\rho]\ket{\vct s}
 =D_{m,n}^{-1}.
 \label{eq:mean-pattern-probability}
\end{equation}
Consequently, for every fixed $A>0$,
\begin{equation}
 \Prob_U\!\left[
 P_{U,\hat\rho}^{(n)}(\vct s)>\frac{m^A}{D_{m,n}}
 \right]\le m^{-A}.
 \label{eq:typical-pattern-probability}
\end{equation}
If $n=\Theta(m)$, then $D_{m,n}=\exp[\Theta(m)]$.
\end{proposition}

\begin{proof}
Irreducibility of $\Sym^n(\C^m)$ and Schur's lemma imply
$\E_U\mathscr C_n(U)[\hat\rho]=\Id/D_{m,n}$, giving
Eq.~\eqref{eq:mean-pattern-probability}.  Markov's inequality gives
Eq.~\eqref{eq:typical-pattern-probability}.  The final statement follows
from Stirling's formula applied to Eq.~\eqref{eq:Dmn}.
\end{proof}

\section{Equal squeezing: exact correlation weights}
\label{sec:equal-squeezing}

When $t_j=t$ for every mode, normalization within a fixed sector removes the
common squeezing magnitude and phase.  Define
\begin{equation}
 \begin{aligned}
 \hat K_+&:=\frac12\sum_{j=1}^m\hat a_j^{\dagger2},
 &\hat K_-&:=\hat K_+^\dagger,\\
 \ket{\Omega_N}
 &:=\frac{\hat K_+^N\ket0}
 {\sqrt{N!\Poch{m/2}{N}}},
 &\hat\rho_{2N}&:=\ket{\Omega_N}\!\bra{\Omega_N}.
 \end{aligned}
 \label{eq:isotropic-state}
\end{equation}
The state is invariant under the real orthogonal subgroup
$O(m)\subset U(m)$.

\begin{proposition}[Exact equal-squeezing coefficients]
\label{prop:equal-Lambda}
For $0\le r<N$,
\begin{equation}
 \Lambda_{\rho,2N,2r+1}=0.
\end{equation}
For $0\le r\le N$,
\begin{align}
 \Lambda_{\rho,2N,2r}
 &=\left[\frac{\Poch{1/2}{r}}{\Poch{m/2}{r}}\right]^2
 \frac{\Poch{-N}{r}}{\Poch{1/2-N}{r}}
 \frac{\Poch{m/2+N}{r}}
      {\Poch{(m+1)/2+N}{r}}.
 \label{eq:equal-Lambda-exact}
\end{align}
For every fixed $r$, uniformly over $N\ge r$,
\begin{equation}
 \Lambda_{\rho,2N,2r}=\Theta_r(m^{-2r}).
 \label{eq:equal-Lambda-scaling}
\end{equation}
\end{proposition}

\begin{proof}
We give the complete conjugation-representation calculation because the
closely related tensor-square overlaps derived in
Ref.~\cite{Kolarovszki2026} use different projectors and cannot be identified
term by term with Eq.~\eqref{eq:Lambda-definition}.

Let $\mathcal A_k:=\mathcal B(\mathcal H_{m,k})$, and let
$\mathcal A_k^{O(m)}$ and $\mathcal P_k^{O(m)}$ be the corresponding
$O(m)$-fixed subspaces of $\mathcal A_k$ and $\mathcal P_k$, respectively.
The
Fischer decomposition~\cite{Howe1989}
\begin{equation}
 \mathcal H_{m,k}
 =\bigoplus_{q=0}^{\lfloor k/2\rfloor}
 \hat K_+^q\mathscr H_{m,k-2q},
 \qquad
 \mathscr H_{m,\ell}:=\ker\hat K_-\cap\mathcal H_{m,\ell},
 \label{eq:Fischer}
\end{equation}
is multiplicity free under $O(m)$.  Hence
$\dim\mathcal A_k^{O(m)}=\lfloor k/2\rfloor+1$.  The orthogonal and
$O(m)$-equivariant splitting
$\mathcal A_k=\mathcal P_k\oplus
\mathsf R_{k-1}(\mathcal A_{k-1})$ therefore gives
\begin{equation}
 \dim\mathcal P_k^{O(m)}
 =\begin{cases}1,&k\text{ even},\\0,&k\text{ odd}.
 \end{cases}
 \label{eq:primitive-invariants}
\end{equation}
Every component of $\hat\rho_{2N}$ is $O(m)$ invariant, proving the odd
selection rule.

For even order $2r$, take the primitive matrix unit
\begin{equation}
 Z_{2r}:=\ket{2r\vct e_1}\!\bra{2r\vct e_2},
 \qquad
 \overline Z_{m,r}:=\int_{O(m)}
 \mathscr C_{2r}(O)[Z_{2r}]\dd O.
 \label{eq:orbit-average}
\end{equation}
It spans the invariant primitive line.  Since group averaging is an
orthogonal projector,
\begin{equation}
 \beta_{m,r}:=\norm{\overline Z_{m,r}}_{\HS}^2
 =\int_{O(m)}O_{11}^{2r}O_{22}^{2r}\dd O.
 \label{eq:beta-integral}
\end{equation}
Condition first on the first column of $O$, use real-sphere moments, and then
apply the terminating Chu--Vandermonde identity.  This gives
\begin{equation}
 \beta_{m,r}
 =\frac{\Poch{1/2}{r}^2}
 {\Poch{(m-1)/2}{r}\Poch{m/2}{r}}
 \frac{\Poch{(m-1)/2+r}{r}}
      {\Poch{m/2+r}{r}}.
 \label{eq:beta-closed}
\end{equation}
For completeness, the sphere moments used here are
\begin{equation}
 \E[x_1^{2p}x_2^{2q}]
 =\frac{\Poch{1/2}{p}\Poch{1/2}{q}}
       {\Poch{m/2}{p+q}},
 \qquad \vct x\sim\mathrm{Unif}(S^{m-1}).
 \label{eq:real-sphere-moments}
\end{equation}

It remains to evaluate one lowering-map matrix element.  The nonzero Fock
coefficients of Eq.~\eqref{eq:isotropic-state} are
\begin{equation}
 \braket{2\vct \ell}{\Omega_N}
 =\frac{\sqrt{N!}}{2^N\sqrt{\Poch{m/2}{N}}}
 \prod_{j=1}^m\frac{\sqrt{(2\ell_j)!}}{\ell_j!},
 \qquad \abs{\vct \ell}=N.
 \label{eq:isotropic-Fock-coefficients}
\end{equation}
Iterating deletion of a Fock matrix unit gives
\begin{align}
 &\mathsf L_{n\to k}(\ket{\vct s}\!\bra{\vct y})\notag\\
 &\quad=(n-k)!
 \sum_{\substack{\vct d\le\vct s,\vct y\\
                   |\vct d|=n-k}}
 \sqrt{\binom{\vct s}{\vct d}
       \binom{\vct y}{\vct d}}
 \ket{\vct s-\vct d}\!\bra{\vct y-\vct d}.
 \label{eq:iterated-deletion}
\end{align}
Consequently, with $a=m/2$,
\begin{align}
 \mu_{m,N,r}
 &:=\left\langle Z_{2r},
 \mathsf L_{2N\to2r}(\hat\rho_{2N})
 \right\rangle_{\HS}\notag\\
 &=(2N-2r)!\frac{N!}{\Poch{a}{N}}
 \frac{\Poch{1/2}{r}}{r!}
 \frac{\Poch{a+2r}{N-r}}{(N-r)!}.
 \label{eq:mu-isotropic}
\end{align}
One verifies Eq.~\eqref{eq:mu-isotropic} by extracting the coefficient of
$z^{N-r}$ from
\begin{equation}
 F_r(z)^2F_0(z)^{m-2},
 \quad
 F_0(z)=(1-4z)^{-1/2},
 \quad
 F_r(z)=4^r\Poch{1/2}{r}(1-4z)^{-r-1/2}.
 \label{eq:isotropic-generating-functions}
\end{equation}

Raise the invariant primitive vector to the $2N$ sector.  Adjointness of
deletion and reinsertion and the ladder norm give
\begin{equation}
 \norm{\hat\rho_{2N,2r}}_{\HS}^2
 =\frac{\mu_{m,N,r}^2}
 {(2N-2r)!\Poch{m+4r}{2N-2r}\beta_{m,r}}.
 \label{eq:isotropic-projected-norm}
\end{equation}
Substitution in Eq.~\eqref{eq:Lambda-definition} is most transparently
performed by taking adjacent ratios.  Direct cancellation yields
\begin{equation}
 \frac{\Lambda_{\rho,2N,2r+2}}
      {\Lambda_{\rho,2N,2r}}
 =\left(\frac{r+1/2}{m/2+r}\right)^2
 \frac{-N+r}{1/2-N+r}
 \frac{m/2+N+r}{m/2+N+r+1/2}.
 \label{eq:equal-Lambda-ratio}
\end{equation}
Since $\Lambda_{\rho,2N,0}=1$, iteration proves
Eq.~\eqref{eq:equal-Lambda-exact}.

Finally, for fixed $r$ and $N\ge r$,
\begin{equation}
 \frac{\Poch{1/2}{r}}{\Poch{m/2}{r}}
 =\prod_{j=0}^{r-1}\frac{2j+1}{m+2j}
 =\Theta_r(m^{-r}),
 \label{eq:sphere-scaling}
\end{equation}
while
$1\le\Poch{-N}{r}/\Poch{1/2-N}{r}\le2^r$ and the final ratio in
Eq.~\eqref{eq:equal-Lambda-exact} is bounded above and below by positive
$r$-dependent constants.  This proves Eq.~\eqref{eq:equal-Lambda-scaling}.
\end{proof}

\begin{lemma}[Uniform equal-squeezing tail]
\label{lem:equal-squeezing-tail}
For every integer $R\ge0$ and $m\ge2$,
\begin{equation}
 \sup_{N\ge r\ge R}\Lambda_{\rho,2N,2r}
 \le
 \frac{\Poch{1/2}{R}R!}{\Poch{m/2}{R}^{2}}
 \le
 4^R\Poch{1/2}{R}R!\,m^{-2R}.
 \label{eq:uniform-equal-squeezing-tail}
\end{equation}
Consequently, for every fixed integer $K\ge0$,
\begin{equation}
 \max_{K<k\le2N}\Lambda_{\rho,2N,k}
 =O_K\!\left(m^{-2(\lfloor K/2\rfloor+1)}\right)
 \label{eq:equal-squeezing-max-tail}
\end{equation}
uniformly over all $N$ with $2N>K$.
\end{lemma}

\begin{proof}
For $N\ge r$,
\begin{equation}
\begin{aligned}
 \frac{\Poch{-N}{r}}{\Poch{1/2-N}{r}}
 &=\prod_{j=0}^{r-1}\frac{N-j}{N-j-1/2}
 \le\prod_{j=1}^{r}\frac{j}{j-1/2}
 =\frac{r!}{\Poch{1/2}{r}},\\
 \frac{\Poch{m/2+N}{r}}{\Poch{(m+1)/2+N}{r}}
 &\le1.
\end{aligned}
 \label{eq:equal-squeezing-ratio-bounds}
\end{equation}
Equation~\eqref{eq:equal-Lambda-exact} therefore gives
\begin{equation}
 \Lambda_{\rho,2N,2r}
 \le B_r
 :=\frac{\Poch{1/2}{r}r!}{\Poch{m/2}{r}^{2}}.
 \label{eq:equal-squeezing-Br}
\end{equation}
For $m\ge2$,
\begin{equation}
 \frac{B_{r+1}}{B_r}
 =\frac{(r+1/2)(r+1)}{(m/2+r)^2}\le1.
 \label{eq:equal-squeezing-Br-ratio}
\end{equation}
Thus $B_r\le B_R$ for $r\ge R$, and
$\Poch{m/2}{R}\ge(m/2)^R$ proves
Eq.~\eqref{eq:uniform-equal-squeezing-tail}.  Taking
$R=\lfloor K/2\rfloor+1$ and using
$\Lambda_{\rho,2N,2r+1}=0$ proves
Eq.~\eqref{eq:equal-squeezing-max-tail}.
\end{proof}

Equation~\eqref{eq:sphere-scaling} is also the geometric origin of the
scaling emphasized in the main text: resolving $r$ specified pair
directions in an isotropic $m$-dimensional mode space costs $m^{-r}$ in
overlap, and the Haar second moment squares that overlap.  Unequal squeezing
magnitudes remove the enlarged $O(m)$ symmetry, so odd orders need not
vanish.

\section{Uniform correlation-order bounds for mode-dependent squeezing}
\label{sec:uniform-correlation-bounds}

The equal-squeezing calculation uses a symmetry unavailable for a general
profile.  We replace it by two complementary estimates.  A state-independent
hierarchy of reduced-state purities controls intermediate and extensive
orders, while a state-specific two-frame calculation controls fixed and
logarithmic orders of the squeezed pair condensate.

\subsection{A multilevel reduced-purity envelope}

For an $n$-boson density operator $\hat\rho$, define its normalized
$\ell$-particle reduced density operator and purity by
\begin{equation}
 \hat\gamma_{\hat\rho}^{(\ell)}
 :=\frac{\ell!}{n!}\mathsf L_{n\to\ell}(\hat\rho),
 \qquad
 \tau_\ell(\hat\rho)
 :=\Tr[(\hat\gamma_{\hat\rho}^{(\ell)})^2]\le1.
 \label{eq:RDM-purity}
\end{equation}
For the squeezed sector we use the shorthand
$\hat\gamma_N^{(\ell)}
:=\hat\gamma_{\hat\rho_{2N}}^{(\ell)}$.
The trace normalization follows from
$\Tr[\mathsf L_q(X)]=q\Tr X$ for $q\ge1$.  Define
\begin{equation}
 G_\ell(\hat\rho)
 :=\norm{\mathsf L_{n\to\ell}(\hat\rho)}_{\HS}^2
 =\left(\frac{n!}{\ell!}\right)^2\tau_\ell(\hat\rho).
 \label{eq:lowered-purity}
\end{equation}
Equation~\eqref{eq:alpha-eigenvalue} and orthogonality give
\begin{equation}
 G_\ell(\hat\rho)
 =\sum_{j=0}^{\ell}
 \alpha_{n,\ell;j}
 \norm{\hat\rho_{n,j}}_{\HS}^2.
 \label{eq:lowered-triangular}
\end{equation}

\begin{lemma}[Multilevel reduced-purity envelope]
\label{lem:multilevel-envelope}
For every $0\le k\le\ell\le n$,
\begin{equation}
 \Lambda_{\rho,n,k}
 \le\mathcal E_{m,n}(k,\ell)\tau_\ell(\hat\rho)
 \le\mathcal E_{m,n}(k,\ell),
 \label{eq:multilevel-envelope}
\end{equation}
where
\begin{equation}
 \mathcal E_{m,n}(k,\ell)
 :=\frac{D_{m,n}}{d_{m,k}}
 \frac{(n!)^2(\ell-k)!}
 {(\ell!)^2(n-k)!\Poch{m+\ell+k}{n-\ell}}.
 \label{eq:envelope-E}
\end{equation}
\end{lemma}

\begin{proof}
The nonnegative $j=k$ term of
Eq.~\eqref{eq:lowered-triangular} implies
$\norm{\hat\rho_{n,k}}_{\HS}^2
\le G_\ell(\hat\rho)/\alpha_{n,\ell;k}$.  Substitute
Eqs.~\eqref{eq:alpha-eigenvalue} and \eqref{eq:lowered-purity}, then multiply
by $D_{m,n}/d_{m,k}$.  The second inequality uses
$\tau_\ell\le1$.
\end{proof}

\begin{lemma}[Optimal lowering level]
\label{lem:optimal-level}
For fixed $m,n,k$,
\begin{equation}
 \frac{\mathcal E_{m,n}(k,\ell+1)}
      {\mathcal E_{m,n}(k,\ell)}
 =\frac{(\ell+1-k)(m+\ell+k)}{(\ell+1)^2}.
 \label{eq:envelope-ratio}
\end{equation}
The sequence decreases and then increases, and a minimizer is
\begin{equation}
 \ell_\star(k)
 =\min\!\left\{n,
 \left\lfloor\frac{k(m+k-1)}{m-1}\right\rfloor
 \right\}.
 \label{eq:optimal-level}
\end{equation}
\end{lemma}

\begin{proof}
Taking the ratio in Eq.~\eqref{eq:envelope-E} leaves
Eq.~\eqref{eq:envelope-ratio}.  That ratio is less than one precisely when
$(\ell+1)(m-1)<k(m+k-1)$, which proves unimodality and
Eq.~\eqref{eq:optimal-level}.
\end{proof}

Two consequences will be used.  First, the endpoint $\ell=k$ gives
\begin{equation}
 \Lambda_{\rho,n,k}
 \le P_{m,n,k}
 :=\frac{\fall nk}{\Poch{m+n}{k}}
   \frac{\Poch{m+k-1}{k}}{\Poch{m-1}{k}}.
 \label{eq:endpoint-envelope}
\end{equation}
Indeed, its product representation implies
\begin{align}
 P_{m,n,k}
 &=\prod_{j=0}^{k-1}
 \frac{n-j}{m+n+j}
 \left(1+\frac{k}{m-1+j}\right)\notag\\
 &\le
 \left(\frac{n}{m+n}\right)^k
 \exp\!\left(\frac{k^2}{m-1}\right).
 \label{eq:sublinear-envelope}
\end{align}
Suppose that $n\le\rho_0m$ and define
\begin{equation}
 \chi_{\rho_0}:=\frac{\rho_0}{1+\rho_0}<1,
 \qquad
 \vartheta:=
 \min\!\left\{\frac14,-\frac14\log\chi_{\rho_0}\right\}.
 \label{eq:middle-order-constants}
\end{equation}
For $m\ge2$ and $k\le\vartheta m$,
$k/(m-1)\le2\vartheta$, and hence
\begin{equation}
 P_{m,n,k}
 \le
 \exp\!\left[
 \bigl(\log\chi_{\rho_0}+2\vartheta\bigr)k
 \right]
 =e^{-c_{\rm mid}k},
 \qquad
 c_{\rm mid}:=
 -\log\chi_{\rho_0}-2\vartheta>0.
 \label{eq:middle-order-bound}
\end{equation}
Thus, for any fixed $A>0$, the correlation weights are exponentially
suppressed throughout the range $A\log m\le k\le\vartheta m$.

Second, optimizing over $\ell$ gives exponential suppression at linear
order.  Appendix~\ref{app:Stirling-envelope} proves the following uniform
form.

\begin{corollary}[State-independent suppression at linear order]
\label{cor:linear-orders}
For every finite $\rho_0$ and every $\vartheta>0$, there are constants
$c_{\vartheta,\rho_0}>0$ and a polynomial
$p_{\vartheta,\rho_0}$ such that, for every $n$-boson state with
$n\le\rho_0m$,
\begin{equation}
 \max_{\vartheta m\le k\le n}
 \Lambda_{\rho,n,k}
 \le p_{\vartheta,\rho_0}(m)e^{-c_{\vartheta,\rho_0}m}.
 \label{eq:linear-order-bound}
\end{equation}
\end{corollary}

\subsection{Primitive two-frame identity}

The universal envelope does not suppress a fixed $k$ for an arbitrary
finite-density bosonic state.  We now use the structure of the squeezed pair
condensate.  For a normalized $n$-boson state define
$\hat\gamma_{\hat\rho}^{(k)}$ by Eq.~\eqref{eq:RDM-purity}.  If
$u,v\in\C^m$ are orthonormal, set
\begin{equation}
 \hat a^\dagger(u):=\sum_{j=1}^m u_j\hat a_j^\dagger,
 \qquad
 \ket{u;k}:=\frac{[\hat a^\dagger(u)]^k}{\sqrt{k!}}\ket0.
 \label{eq:k-condensate}
\end{equation}

\begin{lemma}[Primitive two-frame identity]
\label{lem:primitive-probe}
For every $0\le k\le n$,
\begin{equation}
 \Lambda_{\rho,n,k}
 =A_{m,n,k}\,
 \E_{u\perp v}
 \left|
 \binom nk
 \bra{u;k}\hat\gamma_{\hat\rho}^{(k)}\ket{v;k}
 \right|^2,
 \label{eq:primitive-probe}
\end{equation}
where $(u,v)$ is a Haar-random complex orthonormal two-frame and
\begin{equation}
 A_{m,n,k}:=D_{m,n}\frac{(n-k)!}{\Poch{m+2k}{n-k}}.
 \label{eq:A-mnk}
\end{equation}
When $2k\le n$,
\begin{equation}
 A_{m,n,k}
 =\frac{\Poch m{2k}}
 {\fall nk\,\Poch{m+n}{k}}.
 \label{eq:A-simplified}
\end{equation}
\end{lemma}

\begin{proof}
Take $Z_k=\ket{k\vct e_1}\!\bra{k\vct e_2}\in\mathcal P_k$, which has
unit Hilbert--Schmidt norm, and set
$X_{n,k}=\mathsf R_{k\to n}(Z_k)$.  Iterating the ladder relation gives
\begin{equation}
 \norm{X_{n,k}}_{\HS}^2
 =(n-k)!\Poch{m+2k}{n-k}.
 \label{eq:primitive-chain-norm}
\end{equation}
Schur orthogonality on $\mathcal V_{n,k}$ yields
\begin{equation}
 \E_U\abs{\langle X_{n,k},
 \mathscr C_n(U)[\hat\rho]\rangle_{\HS}}^2
 =\frac{\norm{X_{n,k}}_{\HS}^2}{d_{m,k}}
  \norm{\hat\rho_{n,k}}_{\HS}^2.
 \label{eq:probe-Schur}
\end{equation}
Adjointness and equivariance give
\begin{equation}
 \langle X_{n,k},\mathscr C_n(U)[\hat\rho]\rangle_{\HS}
 =\frac{n!}{k!}
 \bra{u;k}\hat\gamma_{\hat\rho}^{(k)}\ket{v;k},
 \label{eq:probe-RDM}
\end{equation}
where $(u,v)$ is a Haar two-frame. Combining
Eqs.~\eqref{eq:primitive-chain-norm}--\eqref{eq:probe-RDM} gives
\begin{equation}
     \Lambda_{\rho,n,k}
 =\frac{D_{m,n}(n!/k!)^2}
 {(n-k)!\Poch{m+2k}{n-k}}
 \E_{u\perp v}
 \left|\bra{u;k}\hat\gamma_{\hat\rho}^{(k)}\ket{v;k}\right|^2.
\end{equation}
Using $n!/k!=\binom nk(n-k)!$ yields
Eq.~\eqref{eq:primitive-probe}; factorial cancellation gives
Eq.~\eqref{eq:A-simplified}.
\end{proof}

\subsection{Gaussian coefficient extraction}

Write
\begin{equation}
 \mathcal Z(z):=\prod_{j=1}^m(1-zx_j)^{-1/2}
 =\sum_{N\ge0}\mathcal Z_N(\vct x)z^N.
 \label{eq:Z-generating}
\end{equation}
For coefficient extraction, ket and bra amplitudes must be analytically
continued with the same variable $z$.  Define
\begin{align}
 C_z&:=\diag\!\left(\frac{zx_j}{1-zx_j}\right)_{j=1}^m,\notag\\
 A_z&:=-\diag\!\left(\frac{\sqrt z\,t_j}{1-zx_j}\right)_{j=1}^m,
 &
 \widetilde A_z&:=-\diag\!\left(
 \frac{\sqrt z\,\overline t_j}{1-zx_j}\right)_{j=1}^m.
 \label{eq:Gaussian-contraction-matrices}
\end{align}
The square-root branches cancel between paired anomalous contractions.
Importantly, $\widetilde A_z$ is the analytic bra continuation, not the
entrywise conjugate of $A_z$ at nonreal $z$.  Set
\begin{equation}
 \alpha_u(z):=\overline u^{\,T}A_z\overline u,
 \quad
 \beta_v(z):=v^T\widetilde A_zv,
 \quad
 \gamma_{uv}(z):=\overline u^{\,T}C_zv.
 \label{eq:three-contractions}
\end{equation}
Gaussian Wick contraction gives
\begin{equation}
 W_k(z;u,v)
 :=\sum_{p=0}^{\lfloor k/2\rfloor}
 \frac{k!}{(k-2p)!(2^pp!)^2}
 [\alpha_u(z)\beta_v(z)]^p
 \gamma_{uv}(z)^{k-2p}.
 \label{eq:Wick-polynomial}
\end{equation}
Expanding the Gaussian state by pair number then gives the exact identity
\begin{equation}
 \binom{2N}{k}
 \bra{u;k}\hat\gamma_N^{(k)}\ket{v;k}
 =\frac{[z^N]\mathcal Z(z)W_k(z;u,v)}
        {\mathcal Z_N(\vct x)}.
 \label{eq:RDM-Gaussian-coefficient}
\end{equation}

The homogeneity
$\mathcal Z_N(s\vct x)=s^N\mathcal Z_N(\vct x)$ shows that the normalized
sector state is unchanged by $x_j\mapsto sx_j$ and
$t_j\mapsto\sqrt s\,t_j$.  For $N>0$, let $s_N$ be the solution of
$\mu(s_N)=N$ and put $y_j=s_Nx_j$.  Under the gap in
Eq.~\eqref{eq:fugacity-gap}, the tilted variables
$Q_j\sim\mathrm{NB}(1/2,y_j)$ have sum $Q$ with mean $N$ and variance
\begin{equation}
 \sigma^2=\frac12\sum_j\frac{y_j}{(1-y_j)^2},
 \qquad N\le\sigma^2\le N/\delta_{\rm fg}.
 \label{eq:tilted-variance}
\end{equation}
In this subsection, a subscript $\delta$ on a constant abbreviates dependence
only on the fixed fugacity-gap parameter $\delta_{\rm fg}$.

\begin{lemma}[Uniform saddle extraction]
\label{lem:saddle-extraction}
For $N\ge1$, under Eq.~\eqref{eq:fugacity-gap}, there are constants
$c_\delta,C_\delta>0$ such that
\begin{equation}
 p_N(s_N):=\frac{\mathcal Z_Ns_N^N}{\mathcal Z(s_N)}
 \ge\frac{c_\delta}{\sqrt{N+1}},
 \label{eq:local-mass-lower}
\end{equation}
and, for every function $F$ analytic on and inside the saddle circle,
\begin{equation}
 \left|
 \frac{[z^N]\mathcal Z(z)F(z)}{\mathcal Z_N}
 \right|^2
 \le C_\delta\sqrt{N+1}
 \frac1{2\pi}\int_{-\pi}^{\pi}
 \abs{F(s_Ne^{i\theta})}^2\dd\theta.
 \label{eq:saddle-extraction}
\end{equation}
\end{lemma}

The proof, including uniform control for highly nonuniform profiles, is in
Appendix~\ref{app:local-CLT}.

On the saddle circle, the contraction matrices satisfy
\begin{equation}
 \norm{A_z}_{\rm op}+\norm{\widetilde A_z}_{\rm op}
 +\norm{C_z}_{\rm op}\le C_\delta,
 \quad
 \norm{A_z}_{\HS}^2+\norm{\widetilde A_z}_{\HS}^2
 +\norm{C_z}_{\HS}^2\le C_\delta N.
 \label{eq:contraction-norms}
\end{equation}
Indeed, with $z=s_Ne^{i\theta}$ and $y_j=s_Nx_j$, one has
$|1-y_je^{i\theta}|\ge1-y_j\ge\delta_{\rm fg}$ and
$\sum_jy_j/(1-y_j)=2N$.  Thus, for example,
$\norm{A_z}_{\HS}^2\le\sum_jy_j/(1-y_j)^2\le2N/\delta_{\rm fg}$;
the same argument applies to $\widetilde A_z$, while
$\norm{C_z}_{\HS}^2\le\sum_j[y_j/(1-y_j)]^2=O_\delta(N)$.
Since all three matrices are diagonal, the corresponding diagonal-entry
maxima give the operator-norm bounds, uniformly in $\theta$.

\begin{lemma}[Wick moment on a Haar two-frame]
\label{lem:Wick-moment}
For every fixed $A>0$, if $1\le k\le A\log m$, $N\ge k$, and the fugacity gap
holds, then uniformly on the saddle circle,
\begin{equation}
 \E_{u\perp v}\abs{W_k(z;u,v)}^2
 \le\left(\frac{C_{A,\delta}kN}{m^2}\right)^k.
 \label{eq:Wick-moment}
\end{equation}
\end{lemma}

\begin{proof}
For a symmetric matrix $B$, put $S=\norm B_{\HS}^2$ and
$L=\norm B_{\rm op}$.  The exact bounds in
Appendix~\ref{app:Haar-moments} imply
\[
 \E\abs{u^TBu}^{2q}
 \le\left[\frac{4q(S/2+qL^2)}{m^2}\right]^q.
\]
For an arbitrary matrix $C$, with the analogous definitions of $S$ and $L$,
they imply, for $m\ge2$,
\[
 \E_{u\perp v}\abs{u^TC\overline v}^{2q}
 \le\left[\frac{2q(S+qL^2)}{m^2}\right]^q.
\]
By conjugation invariance, the latter estimate also holds for
$\abs{\overline u^{\,T}C v}$.  Equation~\eqref{eq:contraction-norms},
$q\le3k$, and $N\ge k$ therefore give
\begin{equation}
 \max\left\{
 \E\abs{\alpha_u(z)}^{2q},
 \E\abs{\beta_v(z)}^{2q},
 \E_{u\perp v}\abs{\gamma_{uv}(z)}^{2q}
 \right\}
 \le\left(\frac{C_\delta qN}{m^2}\right)^q,
 \qquad 1\le q\le3k.
 \label{eq:simple-Haar-moments}
\end{equation}
For a fixed summand in Eq.~\eqref{eq:Wick-polynomial}, put $r=k-2p$.
H\"older's inequality gives
\[
 \E\abs{\alpha_u^p\beta_v^p\gamma_{uv}^r}^2
 \le
 \bigl(\E\abs{\alpha_u}^{6p}\bigr)^{1/3}
 \bigl(\E\abs{\beta_v}^{6p}\bigr)^{1/3}
 \bigl(\E\abs{\gamma_{uv}}^{6r}\bigr)^{1/3}
 \le\left(\frac{C_{A,\delta}kN}{m^2}\right)^k,
\]
with zero exponents omitted.  Thus the parameter $q$ in
Eq.~\eqref{eq:simple-Haar-moments} never exceeds $3k$ (the corresponding
absolute moment degree never exceeds $6k$).  Finally, weighted
Cauchy--Schwarz and
$c_p=k!/[(k-2p)!(2^pp!)^2]
=\binom{k}{2p}\binom{2p}{p}/4^p\le\binom{k}{2p}$ give
$\sum_pc_p\le2^k$ and absorb the finite sum over $p$.
\end{proof}

\begin{theorem}[Low and logarithmic orders]
\label{thm:low-orders}
Fix $A,\rho_0>0$.  For $N\ge1$, under the fugacity gap, if
$2N\le\rho_0m$, $2N\ge2k$, and $1\le k\le A\log m$, then
\begin{equation}
 \Lambda_{\rho,2N,k}
 \le C\sqrt{N+1}
 \left(\frac{Ck}{m}\right)^k,
 \label{eq:low-order-bound}
\end{equation}
where $C$ depends only on $A,\rho_0$ and $\delta_{\rm fg}$.
\end{theorem}

\begin{proof}
Insert Eqs.~\eqref{eq:RDM-Gaussian-coefficient},
\eqref{eq:saddle-extraction}, and \eqref{eq:Wick-moment} into the primitive
identity \eqref{eq:primitive-probe}.  For $2k\le n=2N$ and
$n\le\rho_0m$, Eq.~\eqref{eq:A-simplified} gives
$A_{m,n,k}\le(C_{\rho_0}m/n)^k$.  This cancels the factor $(N/m^2)^k$
up to $N/n=1/2$ and yields Eq.~\eqref{eq:low-order-bound}.
\end{proof}

\subsection{Uniform fixed-sector tail}

\begin{lemma}[Small photon-number sectors]
\label{lem:small-sectors}
For every fixed nonnegative integer $K$, every fixed $C>0$, and every
$n$-boson state with
$n\le C\log m$,
\begin{equation}
 \max_{K<k\le n}\Lambda_{\rho,n,k}
 \le C_{K,C}\left(\frac{\log m}{m}\right)^{K+1}
 \le C'_{K,C}m^{-K-1/2}.
 \label{eq:small-sector-bound}
\end{equation}
\end{lemma}

\begin{proof}
Equation~\eqref{eq:endpoint-envelope} implies
\begin{equation}
 P_{m,n,k}
 \le\left(\frac{n}{m+n}\right)^k
 \exp\!\left(\frac{k^2}{m-1}\right).
 \end{equation}
For $n\le C\log m$ the exponential factor is at most two for all
$k\le n$, while the base is smaller than $C\log m/m$.  The first omitted
order therefore dominates, and $(\log m)^{K+1}\le m^{1/2}$ for sufficiently
large $m$.
\end{proof}

\begin{theorem}[Uniform fixed-sector correlation tail]
\label{thm:fixed-sector-tail}
Fix an integer $K\ge0$, $B>0$, and $r_{\max}<\infty$.  Let $L_m(B)$ be a physical
cutoff from Lemma~\ref{lem:physical-cutoff}.  There is a constant
$C_{K,B,r_{\max}}$ such that, uniformly over all squeezing profiles obeying
Eq.~\eqref{eq:bounded-squeezing} and all occupied sectors
$0\le N\le L_m(B)$ with $q_N>0$,
\begin{equation}
 \max_{K<k\le2N}\Lambda_{\rho,2N,k}
 \le C_{K,B,r_{\max}}m^{-K-1/2}.
 \label{eq:fixed-sector-tail}
\end{equation}
\end{theorem}

\begin{proof}
Choose $k_0=\lceil A\log m\rceil$, with $A$ fixed below, and set
\begin{equation}
 \rho_0:=\frac{z_1x_\star}{1-z_1x_\star}+1.
 \label{eq:fixed-sector-density-constant}
\end{equation}
Equations~\eqref{eq:physical-cutoff} and \eqref{eq:finite-density} imply
$2N\le\rho_0m$ for every $N\le L_m(B)$ and all sufficiently large $m$.
By
Lemma~\ref{lem:physical-cutoff}, either all relevant sectors have
$2N=O_B(\log m)$, in which case Lemma~\ref{lem:small-sectors} proves the
claim, or every nontrivial relevant sector has the uniform fugacity gap.
Sectors with $2N\le4k_0$ are again covered by
Lemma~\ref{lem:small-sectors}.  For the remaining sectors, we use the squeezed-state bound
for $K<k<k_0$, the endpoint envelope for
$k_0\le k\le\vartheta m$, and the optimized envelope
for $\vartheta m\le k\le2N$.

For $K<k<k_0$, Theorem~\ref{thm:low-orders} gives
\begin{equation}
 \Lambda_{\rho,2N,k}
 \le C\sqrt m\left(\frac{Ck}{m}\right)^k.
 \label{eq:range-one}
\end{equation}
The ratio of the right-hand side at $k+1$ to that at $k$ is $o(1)$ uniformly
in this range, so the first omitted order $k=K+1$ is largest and contributes
$O_K(m^{-K-1/2})$.

For $k_0\le k\le\vartheta m$, Eq.~\eqref{eq:middle-order-bound} gives
\begin{equation}
 \Lambda_{\rho,2N,k}
 \le e^{-c_{\rm mid}k}
 \le m^{-c_{\rm mid}A}.
 \label{eq:middle-range-tail}
\end{equation}
Choose $A$ so that $c_{\rm mid}A\ge K+1$.  Finally, for
$\vartheta m\le k\le2N$ use
Corollary~\ref{cor:linear-orders}; its polynomial times exponential bound is
smaller than $m^{-K-1}$ for sufficiently large $m$.  Combining the ranges
proves Eq.~\eqref{eq:fixed-sector-tail}.  Fixed squeezing phases do not enter
the constants: the contraction bounds use only $|t_j|$, while the exact
algorithm below retains the phases.
\end{proof}

\section{Global truncation in the Haar average case}
\label{sec:global-truncation}

For the bounded outcome function $f_m:\N^m\to[-1,1]$, define the exact global mean
and its deterministic two-cutoff surrogate by
\begin{align}
 \mu_f(U)
 &:=\sum_{N\ge0}q_N\mu_{f,2N}(U),\notag\\
 \mu_f^{(\le K,L_m)}(U)
 &:=\sum_{N=0}^{L_m}q_N
 \mu_{f,2N}^{(\le K)}(U).
 \label{eq:global-bounded-surrogate}
\end{align}
The second expression is a deterministic function of $U$, not the output of
the randomized classical estimator.

\begin{theorem}[Global root-mean-square truncation]
\label{thm:global-RMS}
For every fixed integer $K\ge0$ and every fixed $B>0$,
\begin{equation}
 \norm{\mu_f-\mu_f^{(\le K,L_m)}}_{L^2(U(m))}
 \le C_{K,B,r_{\max}}m^{-K/2-1/4}+m^{-B},
 \label{eq:global-RMS}
\end{equation}
where $L_m=L_m(B)$ may be chosen as in
Lemma~\ref{lem:physical-cutoff}.
\end{theorem}

\begin{proof}
Minkowski's inequality, Theorem~\ref{thm:Schur-identity}, and
Theorem~\ref{thm:fixed-sector-tail} give
\begin{align}
 &\left\|
 \sum_{\substack{N\le L_m\\q_N>0}}q_N
 [\mu_{f,2N}-\mu_{f,2N}^{(\le K)}]
 \right\|_{L^2(U(m))}\notag\\
 &\quad\le\sum_{\substack{N\le L_m\\q_N>0}}q_N
 \sqrt{\max_{K<k\le2N}\Lambda_{\rho,2N,k}}
 \le C_{K,B,r_{\max}}m^{-K/2-1/4}.
 \label{eq:Minkowski-sectors}
\end{align}
This bound requires no independence between photon-number sectors.
Because $|\mu_{f,2N}(U)|\le1$, the omitted physical contribution is bounded
pointwise by $\Prob[N>L_m]\le m^{-B}$.  Adding the two errors proves
Eq.~\eqref{eq:global-RMS}.
\end{proof}

For a target threshold $m^{-c}/2$, Markov's inequality applied to the
squared error gives
\begin{equation}
 \Prob_U\!\left[
 \abs{\mu_f(U)-\mu_f^{(\le K,L_m)}(U)}>\frac12m^{-c}
 \right]
 \le4m^{2c}
 \norm{\mu_f-\mu_f^{(\le K,L_m)}}_{L^2(U(m))}^2.
 \label{eq:Markov-squared-bounded}
\end{equation}
Thus the safe choices
\begin{equation}
 K\ge\lceil2c+\kappa+1\rceil,
 \qquad
 B\ge c+\kappa/2+1
 \label{eq:bounded-cutoffs}
\end{equation}
make the Haar truncation failure at most $m^{-\kappa}/2$ for sufficiently
large $m$, after a harmless adjustment of constants.  No assertion of
exponential decay at every fixed order is used: the first omitted fixed order
is polynomially suppressed, whereas the exponential estimates enter only
when $k$ itself grows with $m$.

\section{Classical evaluation of the retained orders}
\label{sec:classical-estimator}

\subsection{Fixed-order reduced density operators}

For an integer $s\ge0$, define the single-mode sector amplitude
\begin{equation}
 \psi_j(s):=
 \begin{cases}
 t_j^{s/2}\sqrt{\dfrac{\Poch{1/2}{s/2}}{(s/2)!}},
 &s\text{ even},\\[1.2ex]
 0,&s\text{ odd}.
 \end{cases}
 \label{eq:single-mode-amplitude}
\end{equation}
The omitted sign $(-1)^{s/2}$ is common to bra and ket at fixed total pair
number and cancels from the reduced density operator.  For
$|\vct u|=|\vct v|=k$, let
\begin{equation}
 F_{u_j,v_j}^{(j)}(z)
 :=\sum_{\delta\ge0}
 \psi_j(u_j+\delta)\psi_j(v_j+\delta)^*
 \sqrt{\binom{u_j+\delta}{\delta}
       \binom{v_j+\delta}{\delta}}z^\delta.
 \label{eq:F-uv}
\end{equation}

\begin{lemma}[Exact fixed-order reduced state]
\label{lem:RDM-DP}
For an occupied normalized $2N$-photon sector state, $0\le k\le2N$, and
$\vct u,\vct v\in\Phi_{m,k}$,
\begin{equation}
 \bra{\vct u}\hat\gamma_N^{(k)}\ket{\vct v}
 =\frac{[z^{2N-k}]
 \prod_{j=1}^mF_{u_j,v_j}^{(j)}(z)}
 {\binom{2N}{k}\mathcal Z_N}.
 \label{eq:RDM-DP}
\end{equation}
For fixed $k$, all matrix elements can be computed to inverse-polynomial
precision in $m^{O(k)}\poly(N)$ bit operations on every retained sector of
nonnegligible physical weight.
\end{lemma}

\begin{proof}
Write $c_{\vct s}:=\braket{\vct s}{\Omega_N}$.  Iterating particle
deletion gives
\begin{align}
 \bra{\vct u}\hat\gamma_N^{(k)}\ket{\vct v}
 &=\frac1{\binom{2N}{k}}
 \sum_{\substack{\vct\delta\in\N^m\\
 |\vct\delta|=2N-k}}
 c_{\vct u+\vct\delta}
 c_{\vct v+\vct\delta}^*\notag\\
 &\quad\times
 \sqrt{\binom{\vct u+\vct\delta}{\vct\delta}
       \binom{\vct v+\vct\delta}{\vct\delta}}.
 \label{eq:RDM-deletion-sum}
\end{align}
The sector amplitudes factor as
$c_{\vct s}=\mathcal Z_N^{-1/2}\prod_j\psi_j(s_j)$, up to the common sector
phase.  A single generating variable imposes the total-occupation constraint
and gives Eq.~\eqref{eq:RDM-DP}.  There are
$D_{m,k}^2=m^{O(k)}$ entries.  Multiplying the $m$ one-variable series while
retaining coefficients only through $z^{2N-k}$ gives a dynamic program
polynomial in $m,N$ for fixed $k$.
The denominator issue for very rare sectors is handled in
Section~\ref{sec:finite-precision}.
\end{proof}

\subsection{Triangular reconstruction of the retained weights}

For an occupied sector $n=2N$, a pattern $\vct s\in\Phi_{m,2N}$, and
$0\le k\le2N$, define
\begin{equation}
 w_{N,k}(\vct s;U)
 :=\bra{\vct s}
 \mathscr C_{2N}(U)[\hat\rho_{2N,k}]
 \ket{\vct s}.
 \label{eq:wNk}
\end{equation}
For every operator $Y$ on $\mathcal H_{m,k}$, iterating the reinsertion map
gives
\begin{equation}
 \bra{\vct s}\mathsf R_{k\to n}(Y)\ket{\vct s}
 =
 (n-k)!
 \sum_{\substack{\vct r\le\vct s\\|\vct r|=k}}
 \binom{\vct s}{\vct r}
 \bra{\vct r}Y\ket{\vct r}.
 \label{eq:iterated-reinsertion-diagonal}
\end{equation}
Since
$\mathsf L_{n\to k}(\hat\rho_{2N})=(n!/k!)\hat\gamma_N^{(k)}$
and the lowering maps are equivariant, the diagonal matrix element of
$\mathsf D_{2N,k}\mathscr C_{2N}(U)[\hat\rho_{2N}]$ can therefore be
expressed as
\begin{align}
 b_{N,k}(\vct s;U)
 &:=\frac{(n-k)!n!}{k!}
 \sum_{\substack{\vct r\le\vct s\\|\vct r|=k}}
 \binom{\vct s}{\vct r}
 \bra{\vct r}
 \hat{\mathcal U}^{(k)}(U)
 \hat\gamma_N^{(k)}
 \hat{\mathcal U}^{(k)}(U)^\dagger
 \ket{\vct r}.
 \label{eq:bNk}
\end{align}
Taking diagonal elements of the triangular projector relation gives
\begin{equation}
 w_{N,k}(\vct s;U)
 =\frac{b_{N,k}(\vct s;U)-\sum_{j=0}^{k-1}
 \alpha_{n,k;j}w_{N,j}(\vct s;U)}
 {(n-k)!\Poch{m+2k}{n-k}}.
 \label{eq:w-recursion}
\end{equation}
The required fixed-particle matrix elements are
\begin{equation}
 \bra{\vct r}\hat{\mathcal U}^{(k)}(U)\ket{\vct u}
 =\frac{\Per(U[\vct r,\vct u])}
 {\sqrt{\vct r!\vct u!}},
 \label{eq:permanent-matrix-elements}
\end{equation}
where rows and columns are repeated according to the occupation vectors
\cite{Scheel2004}.  All permanents have size at most the retained order
$K$, so Eqs.~\eqref{eq:RDM-DP}--\eqref{eq:w-recursion} require
$m^{O(K)}\poly(N)$ time.

\subsection{Unbiased estimator and second moment}

Define
\begin{equation}
 W_{N,K}(\vct s;U)
 :=\sum_{k=0}^{\min\{K,2N\}}w_{N,k}(\vct s;U).
 \label{eq:retained-weight}
\end{equation}
Because $\mathscr C_{2N}(U)[\hat\rho_{2N,k}]\in\mathcal V_{2N,k}$ and
$\mathcal Q_{2N,k}$ is self-adjoint,
\begin{align}
 \sum_{\vct s\in\Phi_{m,2N}}
 f_m(\vct s)w_{N,k}(\vct s;U)
 &=\Tr\!\left[
 \hat O_{f,2N}\mathscr C_{2N}(U)[\hat\rho_{2N,k}]
 \right]\notag\\
 &=\Tr\!\left[
 \hat O_{f,2N,k}\mathscr C_{2N}(U)[\hat\rho_{2N,k}]
 \right]
 =\mu_{f,2N}^{(k)}(U).
 \label{eq:weight-contribution-identity}
\end{align}
Consequently,
\begin{equation}
 \sum_{\vct s\in\Phi_{m,2N}}
 f_m(\vct s)W_{N,K}(\vct s;U)
 =\mu_{f,2N}^{(\le K)}(U).
 \label{eq:retained-weight-identity}
\end{equation}
One ideal Monte Carlo draw for the bounded-function surrogate is:
\begin{enumerate}
 \item draw $N$ from $q_N$ and output zero if $N>L_m$;
 \item draw $\vct S$ uniformly from $\Phi_{m,2N}$;
 \item evaluate $f_m(\vct S)$ once;
 \item compute $W_{N,K}(\vct S;U)$ and output
 \begin{equation}
  Z_f:=D_{m,2N}f_m(\vct S)W_{N,K}(\vct S;U).
  \label{eq:Zf}
 \end{equation}
\end{enumerate}
Uniform composition sampling gives
\begin{equation}
 \E[Z_f\mid N,U]
 =\mathbf1_{\{N\le L_m\}}\mu_{f,2N}^{(\le K)}(U),
 \qquad
 \E Z_f=\mu_f^{(\le K,L_m)}(U).
 \label{eq:Zf-unbiased}
\end{equation}

Let
\begin{equation}
 \hat\sigma_{N,K}(U)
 :=\sum_{k=0}^{\min\{K,2N\}}
 \mathscr C_{2N}(U)[\hat\rho_{2N,k}].
 \label{eq:sigma-truncated}
\end{equation}
Since diagonal squared norm is at most Hilbert--Schmidt squared norm,
\begin{align}
 \E[|Z_f|^2\mid N,U]
 &\le D_{m,2N}\norm{\hat\sigma_{N,K}(U)}_{\HS}^2\notag\\
 &=\sum_{k=0}^{\min\{K,2N\}}
 d_{m,k}\Lambda_{\rho,2N,k}.
 \label{eq:Zf-second-moment}
\end{align}
The scalar component has $\Lambda_{\rho,2N,0}=1$.  For each fixed
$1\le k\le K$, apply Theorem~\ref{thm:fixed-sector-tail} with cutoff $k-1$;
together with $d_{m,k}=\Theta_k(m^{2k})$, this shows that the right-hand side of
Eq.~\eqref{eq:Zf-second-moment} is polynomial in $m$, uniformly for
occupied $N\le L_m$ and all $U$.  More explicitly, after increasing the
constructive constants in Theorem~\ref{thm:fixed-sector-tail} to cover the
finitely many small values of $m$, define
\begin{equation}
 \Sigma_{m,K}
 :=1+\sum_{k=1}^K
 d_{m,k}C_{k-1,B,r_{\max}}m^{-k+1/2}.
 \label{eq:explicit-Sigma}
\end{equation}
For fixed $K,B,r_{\max}$, the constants are hard-coded from the proof,
$\Sigma_{m,K}$ is a computable polynomial envelope, and
the right-hand side of Eq.~\eqref{eq:Zf-second-moment} is at most
$\Sigma_{m,K}$.  Moreover, $d_{m,k}=\Theta_k(m^{2k})$ gives
\begin{equation}
 \Sigma_{m,K}
 =O_{K,B,r_{\max}}\!\left(m^{K+1/2}\right).
 \label{eq:Sigma-polynomial-bound}
\end{equation}
Thus, for every occupied $N\le L_m(B)$ and every $U$,
\begin{equation}
 \E[|Z_f|^2\mid N,U]
 =O_{K,B,r_{\max}}\!\left(m^{K+1/2}\right).
 \label{eq:Zf-second-moment-polynomial}
\end{equation}
A median-of-means estimator therefore estimates
$\mu_f^{(\le K,L_m)}(U)$ to inverse-polynomial additive error with
inverse-polynomial failure probability using polynomially many ideal draws.

The operator $\hat\sigma_{N,K}$ need not be positive, and
$W_{N,K}$ may be signed.  Positivity is neither asserted nor needed.  The
ideal estimator retains every occupation pattern within every sector
$N\le L_m$ and truncates only correlation order.  The finite-precision
implementation below may omit sectors of inverse-polynomially small total
physical weight.  In either case the construction estimates one requested
expectation, not a probability model or sampler for the GBS distribution.

Combining Theorem~\ref{thm:global-RMS}, the second-moment estimate above,
and the finite-precision implementation in
Section~\ref{sec:finite-precision} gives the bounded result that will be used
as a black box below.

\begin{theorem}[Bounded outcome functions]
\label{thm:bounded-functions}
Let $f_m:\N^m\to[-1,1]$ be efficiently evaluable and independent of $U$.
Let $0<\varepsilon_m,\delta_m<1$ be known scales satisfying
$\varepsilon_m^{-1},\delta_m^{-1}\le m^{O(1)}$.  There is a randomized
classical polynomial-time algorithm
that outputs $\widetilde\mu_f(U)$ and satisfies
\begin{equation}
 \Prob_{U\sim\Haar,\,\mathcal A}
 \left[|\widetilde\mu_f(U)-\mu_f(U)|>\varepsilon_m\right]
 \le\delta_m.
 \label{eq:bounded-function-guarantee}
\end{equation}
The polynomial exponent may depend on fixed polynomial bounds for
$\varepsilon_m^{-1}$ and $\delta_m^{-1}$.
\end{theorem}

\begin{proof}
Choose fixed exponents $c_{\rm acc},\kappa_0>0$ so that, for all sufficiently
large $m$, $m^{-c_{\rm acc}}\le\varepsilon_m/4$ and
$m^{-\kappa_0}\le\delta_m/2$.  Apply Eq.~\eqref{eq:bounded-cutoffs} with
$c=c_{\rm acc}$ and $\kappa=\kappa_0$.  It and the restricted form of
Eq.~\eqref{eq:Minkowski-sectors} ensure that the absolute
correlation-error term in Eq.~\eqref{eq:retained-error-decomposition} exceeds
$\varepsilon_m/4$ only on a set of Haar measure at most $\delta_m/2$.  Increase
$B$ by a fixed amount so that $m^{-B}\le\varepsilon_m/16$, and use
$\varepsilon=\varepsilon_m$ in Eq.~\eqref{eq:rare-sector-threshold}; the two
pointwise terms in Eq.~\eqref{eq:retained-error-decomposition} then sum to
at most $3\varepsilon_m/32$.  Conditional on the supplied $U$, the
finite-precision median-of-means construction estimates the retained
surrogate to error at most $\varepsilon_m/2$, with algorithmic failure
probability at most $\delta_m/2$.  These three error budgets sum to less
than $\varepsilon_m$.  The second-moment bound
Eq.~\eqref{eq:Zf-second-moment} and the finite-precision construction in
Section~\ref{sec:finite-precision} make both its sample count and the cost of
each sample polynomial.  A union bound proves
Eq.~\eqref{eq:bounded-function-guarantee}.
\end{proof}

\section{Extension to the general outcome-function class}
\label{sec:general-functions}

Fix a target additive error
\begin{equation}
 \varepsilon:=m^{-c},
 \label{eq:general-target-error}
\end{equation}
and choose the bounds in Eqs.~\eqref{eq:mean-variance-scales}--
\eqref{eq:Haar-good-set-probability} with $\lambda=\kappa+2$.  The phase
reconstruction bounds below are deterministic once $U$ lies in the resulting
good set and all bounded-estimation subroutines succeed.  Write
\begin{equation}
 \begin{aligned}
 F&:=f_m(\vct S),
 &\mu&:=\E_{\vct S\sim P_U}F=\mu_f(U),\\
 M&:=M_m,
 &V&:=V_m.
 \end{aligned}
 \label{eq:F-mu-M-V}
\end{equation}
Then $|\mu|\le M$ and $\operatorname{Var}(F)\le V$.

Choose a polynomial local scale by setting
\begin{equation}
 R:=2^{\left\lceil
 \log_2\!\left(1024\max\left\{1,\sqrt V,\frac V\varepsilon\right\}
 \right)\right\rceil}.
 \label{eq:phase-local-scale}
\end{equation}
For $j=0,1,\ldots,J$, let
\begin{equation}
 T_j:=2^jR,
 \qquad
 J:=\min\{j\ge0:T_j\ge8M\},
 \qquad
 \eta:=\frac{\varepsilon}{1024R}.
 \label{eq:phase-scales}
\end{equation}
Because $V=\poly(m)$ and $\log M=\poly(m)$, both $R$ and $J$ are
polynomially bounded, and $\eta^{-1}=\poly(m)$.

For every scale define two outcome functions derived from the same $f_m$,
\begin{equation}
 \begin{aligned}
 f_{m,j}^{(\mathrm c)}(\vct s)
 &:=\cos\!\left(\frac{2\pi f_m(\vct s)}{T_j}\right),\\
 f_{m,j}^{(\mathrm s)}(\vct s)
 &:=\sin\!\left(\frac{2\pi f_m(\vct s)}{T_j}\right).
 \end{aligned}
 \label{eq:bounded-characteristic-scores}
\end{equation}
They take values in $[-1,1]$, are fixed independently of $U$, and are
efficiently evaluable.  Thus Theorem~\ref{thm:bounded-functions} applies to
all $2(J+1)$ of them.

\begin{lemma}[Variance control of the characteristic function]
\label{lem:characteristic-variance}
For any real random variable $F$ with mean $\mu$ and variance at most $V$,
\begin{equation}
 \left|\E e^{i\omega F}-e^{i\omega\mu}\right|
 \le\frac{\omega^2V}{2}
 \label{eq:characteristic-variance}
\end{equation}
for every $\omega\in\R$.
\end{lemma}

\begin{proof}
Let $Y:=F-\mu$.  Then $\E Y=0$ and $\E Y^2\le V$.  The elementary bound
\begin{equation}
 |e^{ix}-1-ix|\le\frac{x^2}{2}
 \label{eq:exponential-remainder}
\end{equation}
gives
\begin{align}
 \left|\E e^{i\omega Y}-1\right|
 &=\left|\E(e^{i\omega Y}-1-i\omega Y)\right|\notag\\
 &\le\frac{\omega^2}{2}\E Y^2
 \le\frac{\omega^2V}{2}.
\end{align}
Multiplication by $e^{i\omega\mu}$ proves
Eq.~\eqref{eq:characteristic-variance}.
\end{proof}

Set
\begin{equation}
 z_j:=\E e^{2\pi iF/T_j}.
 \label{eq:exact-characteristic-values}
\end{equation}
By Lemma~\ref{lem:characteristic-variance},
\begin{equation}
 \left|z_j-e^{2\pi i\mu/T_j}\right|
 \le\frac{2\pi^2V}{T_j^2}.
 \label{eq:characteristic-phase-bias}
\end{equation}
Use Theorem~\ref{thm:bounded-functions} to obtain estimates
$\widehat c_j$ and $\widehat s_j$ of the expectations of the two functions
in Eq.~\eqref{eq:bounded-characteristic-scores}, each to additive error
$\eta$, and define
\begin{equation}
 \widehat z_j:=\widehat c_j+i\widehat s_j.
 \label{eq:estimated-characteristic-values}
\end{equation}
Whenever all these bounded estimates succeed,
\begin{equation}
 \left|\widehat z_j-e^{2\pi i\mu/T_j}\right|
 \le\rho_j,
 \qquad
 \rho_j:=\frac{2\pi^2V}{T_j^2}+\sqrt2\,\eta<\frac12.
 \label{eq:characteristic-total-error}
\end{equation}

\begin{lemma}[Stable phase and unique dyadic lift]
\label{lem:phase-unwrapping}
Let $\operatorname{Arg}$ take values in $[-\pi,\pi)$ and define the principal
residue
\begin{equation}
 \widetilde\mu_j:=\frac{T_j}{2\pi}\operatorname{Arg}(\widehat z_j)
 \in[-T_j/2,T_j/2).
 \label{eq:phase-residue}
\end{equation}
Then its circular distance from $\mu$ satisfies
\begin{equation}
 d_{T_j}(\widetilde\mu_j,\mu)
 :=\min_{q\in\mathbb Z}|\widetilde\mu_j-\mu-qT_j|
 \le b_j,
 \qquad
 b_j:=\frac{2\pi V}{T_j}+\frac{\sqrt2}{\pi}T_j\eta,
 \label{eq:phase-residue-error}
\end{equation}
and $b_j/T_j<1/64$.  Taking
$\widetilde\mu_J\in[-T_J/2,T_J/2)$, set
\begin{equation}
 \widehat\mu_J:=\widetilde\mu_J,
 \qquad
 \widehat\mu_j:=\widetilde\mu_j+T_j\operatorname{nint}\!\left(
 \frac{\widehat\mu_{j+1}-\widetilde\mu_j}{T_j}\right),
 \qquad j=J-1,\ldots,0,
 \label{eq:dyadic-unwrapping}
\end{equation}
where $\operatorname{nint}$ denotes the nearest integer.  Then
$|\widehat\mu_j-\mu|\le b_j$ for every $j$.
\end{lemma}

\begin{proof}
If $|w-e^{i\theta}|\le\rho<1/2$, rotating by $e^{-i\theta}$ shows that
the phase error is at most
$\arctan[\rho/(1-\rho)]\le2\rho$.  Applying this to
Eq.~\eqref{eq:characteristic-total-error} and multiplying the angular error
by $T_j/(2\pi)$ gives Eq.~\eqref{eq:phase-residue-error}.  The definitions
of $R$ and $\eta$ imply
\begin{equation}
 \frac{b_j}{T_j}
 \le\frac{2\pi V}{R^2}+\frac{\sqrt2}{\pi}\eta<\frac1{64}.
 \label{eq:relative-phase-error}
\end{equation}
Since $T_J\ge8M$ and $|\mu|\le M$, the principal representative
$\widetilde\mu_J$ is the unique lift within $b_J$ of $\mu$.  More explicitly,
$|\mu|\le T_J/8$ and $b_J<T_J/64$ imply that no nonzero translate of
$\widetilde\mu_J\in[-T_J/2,T_J/2)$ can be within $b_J$ of $\mu$; hence
$|\widetilde\mu_J-\mu|\le b_J$.  Suppose inductively that
$|\widehat\mu_{j+1}-\mu|<T_{j+1}/64=T_j/32$.  The correct lift of
$\widetilde\mu_j$ is within
$T_j/64$ of $\mu$ and hence within $3T_j/64$ of $\widehat\mu_{j+1}$, whereas every
other lift differs by at least $61T_j/64$.  The nearest-integer rule in
Eq.~\eqref{eq:dyadic-unwrapping} therefore selects the correct lift and
resets the error to at most $b_j$.
\end{proof}

The algorithm returns $\widehat\mu_0$.  At the finest scale,
Lemma~\ref{lem:phase-unwrapping} gives
\begin{align}
 |\widehat\mu_0-\mu|
 &\le\frac{2\pi V}{R}+\frac{\sqrt2}{\pi}R\eta\notag\\
 &\le\left(\frac{2\pi}{1024}
 +\frac{\sqrt2}{1024\pi}\right)\varepsilon
 <0.007\varepsilon<\varepsilon.
 \label{eq:general-final-error}
\end{align}

\begin{proof}[Proof of Theorem~\ref{thm:main-supp}]
There are $Q_m:=2(J+1)=\poly(m)$ bounded functions in
Eq.~\eqref{eq:bounded-characteristic-scores}.  Apply
Theorem~\ref{thm:bounded-functions} to each one with accuracy $\eta$ and
failure probability at most $m^{-(\kappa+2)}/Q_m$.  A union bound makes the
probability that any bounded estimate fails at most $m^{-(\kappa+2)}$.
Separately, Eq.~\eqref{eq:Haar-good-set-probability} with
$\lambda=\kappa+2$ bounds the probability that $U$ lies outside the set on
which $|\mu|\le M$ and $\operatorname{Var}(F)\le V$ by
$m^{-(\kappa+2)}$.  On the complementary event,
Eq.~\eqref{eq:general-final-error} proves the desired additive accuracy.
For all sufficiently large $m$, the sum of these two failure probabilities
is at most $m^{-\kappa}$.

The number of bounded-estimation calls is polynomial, and each
requires only inverse-polynomial statistical accuracy $\eta$;
phase unwrapping uses polynomial-bit deterministic arithmetic.  Although
$T_j$ and the lift integers may have exponentially large numerical values,
their bit lengths are polynomial because $\log T_j=\poly(m)$.
Polynomial-bit evaluation of $f_m$ therefore permits reduction of its value
modulo $T_j$ and evaluation of the trigonometric functions in polynomial
time.  The detailed bit-level
implementation is given in Section~\ref{sec:finite-precision}.
\end{proof}

\begin{remark}
The physical variance bound in Eq.~\eqref{eq:Haar-good-set} does not control
the second moment of the signed, uniform-proposal estimator obtained by
inserting a general $f_m$ directly into Eq.~\eqref{eq:Zf}.  The classical
Monte Carlo estimator is therefore applied only to the bounded functions in
Eq.~\eqref{eq:bounded-characteristic-scores}; the mean of the original
function is recovered from their phases.
\end{remark}

\section{Predefined readout families and the boundary of the proof}
\label{sec:readout-families}

\subsection{A polynomial-size family fixed in advance}

The independence condition can be relaxed from one fixed readout to a
polynomial-size list fixed before the interferometer is drawn.

\begin{proposition}[Simultaneous estimation for a predefined family]
\label{prop:dictionary}
For each $m$, let
\begin{equation}
 \mathfrak F_m=\{f_{\ell,m}\}_{\ell=1}^{R_m},
 \qquad R_m\le m^\xi,
 \label{eq:dictionary}
\end{equation}
be fixed independently of $U$, where $\xi$ is fixed and all members are
uniformly efficiently evaluable, have polynomial-bit outputs, and share
uniform polynomial bounds on these costs.  Suppose that, for every fixed
$\lambda>0$, there are common known bounds $M_m(\lambda)$ and
$V_m(\lambda)$ satisfying Eq.~\eqref{eq:mean-variance-scales} and, uniformly
for $1\le\ell\le R_m$,
\begin{equation}
\begin{gathered}
\mu_\ell(U)
:=
\E_{\vct S\sim P_U}
[f_{\ell,m}(\vct S)],
\\[2pt]
\Prob_{U\sim\Haar}\!\left[
|\mu_\ell(U)|>M_m
\ \text{or}\ 
\operatorname{Var}_{\vct S\sim P_U}
[f_{\ell,m}(\vct S)]>V_m
\right]
\le m^{-\lambda}.
\end{gathered}
\label{eq:dictionary-good-sets}
\end{equation}
Then, for every fixed $c,\kappa>0$ and all sufficiently large $m$,
polynomial time suffices to produce
estimates $\widetilde\mu_{\ell}(U)$ such that
\begin{equation}
 \Prob_{U\sim\Haar,\,\mathcal A}\!\left[
 \max_{1\le\ell\le R_m}
 |\widetilde\mu_{\ell}(U)-\mu_{\ell}(U)|>m^{-c}
 \right]
 \le m^{-\kappa}.
 \label{eq:dictionary-simultaneous}
\end{equation}
Consequently, after the estimates have been formed, a polynomial-time rule
may select any member of the predefined family as a function of $U$ and of
the estimates themselves.
\end{proposition}

\begin{proof}
Apply Theorem~\ref{thm:main-supp} to every $f_{\ell,m}$ with the same target
accuracy $m^{-c}$ and with failure exponent $\kappa+\xi+1$.  A union bound
gives
\begin{equation}
 \Prob\!\left[
 \max_{1\le\ell\le R_m}
 |\widetilde\mu_{\ell}-\mu_{\ell}|>m^{-c}
 \right]
 \le R_m m^{-(\kappa+\xi+1)}
 \le m^{-(\kappa+1)}.
 \label{eq:dictionary-union}
\end{equation}
The harmless slack absorbs the finitely many small values of $m$.  Running
at most $m^\xi$ polynomial-time estimators remains polynomial time.  The
common success event holds for every member simultaneously, so subsequent
selection from this list introduces no new approximation error.
\end{proof}

\subsection{Why arbitrary interferometer-dependent readout is different}

For a general readout $f_{m,U}$, an order-$k$ contribution contains
\begin{equation}
 \left\langle
 \hat O_{f_{m,U},2N,k},
 \mathscr C_{2N}(U)[\hat\rho_{2N,k}]
 \right\rangle_{\HS}.
 \label{eq:U-correlated-readout}
\end{equation}
Here $\hat O_{f_{m,U},2N,k}$ is the order-$k$ component of the sector operator
induced by $f_{m,U}|_{\Phi_{m,2N}}$.  It is correlated with the same $U$ that
rotates the state, so the Schur separation in
Eq.~\eqref{eq:Schur-four-vector} no longer applies.  For a worst-case
interferometer, the Haar average and its concentration step are absent as
well.  Proposition~\ref{prop:dictionary} works because only polynomially
many readouts, all fixed before $U$ is drawn, need to be controlled on one
common event; it does not cover an arbitrary efficiently computable
$U$-dependent outcome function.

The result also assumes a system-size-independent bound on per-mode
squeezing and the standard passive, lossless architecture.  It does not
cover growing squeezing, displacement, general loss, adaptive optical
measurements, worst-case structured interferometers, or multiplicative
estimation of exponentially small output probabilities.  These cases lie
outside the theorem stated in the main text.

\section{Finite-precision polynomial-time implementation}
\label{sec:finite-precision}

The estimators constructed above are formulated using exact sampling and
arithmetic. To complete the complexity claim, we now show that they can be
implemented as randomized polynomial-time algorithms in the standard bit
model. The required discrete samples can be generated to
inverse-polynomial accuracy, sufficiently low-probability sectors can be discarded so
that all remaining divisions are well conditioned, and the resulting signed
estimators can be evaluated and averaged using only polynomially many bits,
operations, and samples. We first establish this for
$f_m:\N^m\to[-1,1]$ and then verify that the characteristic-function
reduction for general $f_m$ preserves the same complexity.

\subsection{Discrete sampling and well-conditioned sectors}

The real number $\mu(z_1)$ in Eq.~\eqref{eq:physical-cutoff} need not have an
exact finite encoding.  Certified interval arithmetic can produce a rational
upper bound $\overline\mu(z_1)$ with
$\mu(z_1)\le\overline\mu(z_1)\le\mu(z_1)+1$.  Operationally, we use the
integer obtained from Eq.~\eqref{eq:physical-cutoff} with
$\overline\mu(z_1)$ in place of $\mu(z_1)$.  It lies between $L_m(B)$ and
$L_m(B)+1$; the pair-number tail is unchanged, and the fugacity-gap proof
remains valid after increasing the fixed slack $H_m$ by one.  We retain the
notation $L_m$ for this certified cutoff.

The physical pair number can be sampled without evaluating a hafnian. The
independent single-mode variables in Eq.~\eqref{eq:single-mode-NB} obey
\begin{equation}
 \frac{\Prob[N_j=q+1]}{\Prob[N_j=q]}
 =x_j\frac{q+1/2}{q+1}.
 \label{eq:NB-recurrence}
\end{equation}
For a target sampling error $0<\xi_{\rm samp}<1$, because
$x_j\le x_\star<1$,
truncating each recurrence after
$O(\!\log(m/\xi_{\rm samp}))$ terms makes the total overflow probability at
most $\xi_{\rm samp}/4$.  Certified interval arithmetic then produces dyadic
approximations to the truncated single-mode laws whose product law is within
$\xi_{\rm samp}/4$ in
total variation; fixed-bit inversion sampling from these dyadic laws is
exact.  Hence the resulting $N=\sum_jN_j$ is within total-variation distance
$\xi_{\rm samp}/2$ of its target law in worst-case polynomial time.
Alternatively, all
$q_N$ with $N\le L_m=O(m)$ can be
obtained by truncated convolution. Conditional on $N$, stars and bars
identifies $\Phi_{m,2N}$ with the
$\binom{2N+m-1}{m-1}$ subsets of $m-1$ bar positions. Drawing a uniform
subset rank by rejection from
$\lceil\log_2\binom{2N+m-1}{m-1}\rceil$ fair bits gives an exact sample in
expected polynomial time, with acceptance probability greater than $1/2$.
For worst-case polynomial runtime, cap the rejection procedure after
$O(\log(1/\xi_{\rm samp}))$ attempts and return an arbitrary pattern on
failure.  The resulting conditional law is within $\xi_{\rm samp}/2$ in total
variation, so the combined law of $(N,\vct S)$ is within $\xi_{\rm samp}$ of
its target law.

It remains to control divisions by $\mathcal Z_N(\vct x)$, which can be
exponentially small for a highly nonuniform squeezing profile. For a target
error $\varepsilon>0$, compute each $q_N$ with certified absolute error at
most $\tau$ and discard every sector satisfying
\begin{equation}
 \widetilde q_N<2\tau,
 \qquad
 \tau:=\frac{\varepsilon}{96(L_m+1)},
 \label{eq:rare-sector-threshold}
\end{equation}
and define the retained set and its deterministic surrogate by
\begin{align}
 \mathcal R_m
 &:=\{0\le N\le L_m:\ \widetilde q_N\ge2\tau\},
 \notag\\
 \mu_f^{(\le K,\mathcal R_m)}(U)
 &:=\sum_{N\in\mathcal R_m}q_N
 \mu_{f,2N}^{(\le K)}(U).
 \label{eq:retained-sector-surrogate}
\end{align}
Every discarded sector has $q_N<3\tau$, and hence
\begin{equation}
 \sum_{\substack{0\le N\le L_m\\N\notin\mathcal R_m}}q_N
 \le3(L_m+1)\tau=\varepsilon/32.
 \label{eq:discarded-sector-mass}
\end{equation}
Since $|f_m|\le1$, the same bound applies to the omitted contribution to the
mean. More precisely,
\begin{align}
 \mu_f(U)-\mu_f^{(\le K,\mathcal R_m)}(U)
 &=\sum_{N>L_m}q_N\mu_{f,2N}(U)
 \notag\\
 &\quad+\sum_{\substack{0\le N\le L_m\\N\notin\mathcal R_m}}
 q_N\mu_{f,2N}(U)
 \notag\\
 &\quad+\sum_{N\in\mathcal R_m}q_N
 \left[
 \mu_{f,2N}(U)-\mu_{f,2N}^{(\le K)}(U)
 \right].
 \label{eq:retained-error-decomposition}
\end{align}
The first two terms are deterministically bounded by
$m^{-B}+\varepsilon/32$. The argument leading to
Eq.~\eqref{eq:Minkowski-sectors} bounds the $L^2(U(m))$ norm of the last term
by $C_{K,B,r_{\max}}m^{-K/2-1/4}$; only this correlation-truncation term
requires Markov's inequality.

For $N\in\mathcal R_m$, certified approximation gives
\begin{equation}
 \mathcal Z_N(\vct x)\ge q_N\ge\tau,
 \label{eq:ZN-lower-bound}
\end{equation}
where the first inequality follows from Eq.~\eqref{eq:pair-law}. Thus every
division in Eq.~\eqref{eq:RDM-DP} is by an inverse-polynomially bounded
quantity. For fixed $K$ and $N\le L_m$, all factorials and Pochhammer
symbols have polynomial bit length, the coefficient-extraction programs
have polynomial length, and the permanents in
Eq.~\eqref{eq:permanent-matrix-elements} have fixed size. Standard
multiprecision arithmetic therefore evaluates each retained weight to the
required accuracy using $\poly(m,\log(1/\varepsilon))$ bit operations
\cite{BrentZimmermann2010}.

The accuracy budget applies to the complete sample in Eq.~\eqref{eq:Zf}, not
to $W_{N,K}$ alone.  Put
$D=D_{m,2N}$.  Equations~\eqref{eq:Zf-second-moment} and
\eqref{eq:explicit-Sigma} imply, pointwise in every retained pattern,
\[
 |W_{N,K}(\vct S;U)|
 \le\norm{\hat\sigma_{N,K}(U)}_{\HS}
 \le\sqrt{\frac{\Sigma_{m,K}}{D}},
 \qquad
 |D W_{N,K}(\vct S;U)|\le\sqrt{D\Sigma_{m,K}}.
\]
For a desired per-sample error $0<\eta_{\rm samp}\le1$, it is sufficient,
for example, to compute $W_{N,K}$ to absolute error
$\eta_{\rm samp}/(4D)$ and $f_m(\vct S)$ to absolute error
\[
 \frac{\eta_{\rm samp}}
 {4\bigl(D+\sqrt{D\Sigma_{m,K}}\bigr)},
\]
with the remaining constant fraction reserved for the final multiplication.
The resulting guard-bit budget is
$O(\log D+\log(1+\Sigma_{m,K})+\log\eta_{\rm samp}^{-1})=\poly(m)$ because
$N\le L_m=O(m)$.  Below we take $\eta_{\rm samp}=\varepsilon/64$.

\subsection{Finite-precision estimation for bounded functions}

For a draw $N\sim q_N$, let
\begin{equation}
 Z_f^{\mathrm{ret}}
 :=\mathbf1_{\{N\in\mathcal R_m\}}Z_f.
 \label{eq:retained-Zf}
\end{equation}
Equations~\eqref{eq:Zf-unbiased} and
\eqref{eq:retained-sector-surrogate} give
\begin{equation}
 \E Z_f^{\mathrm{ret}}
 =\mu_f^{(\le K,\mathcal R_m)}(U).
 \label{eq:retained-Zf-unbiased}
\end{equation}
Although this variable need not be pointwise polynomially bounded,
Eqs.~\eqref{eq:Zf-second-moment} and \eqref{eq:explicit-Sigma} give
\begin{equation}
 \E[|Z_f^{\mathrm{ret}}|^2\mid N,U]
 \le\Sigma_{m,K}
 \label{eq:Sigma-bound}
\end{equation}
for every $N$ and $U$, where $\Sigma_{m,K}=\poly(m)$ for fixed $K$.
Clip the real output at
\begin{equation}
 B_{\mathrm{clip}}
 :=\frac{32\Sigma_{m,K}}{\varepsilon},
 \qquad
 \operatorname{clip}_{B_{\mathrm{clip}}}(z)
 :=\max\{-B_{\mathrm{clip}},\min\{z,B_{\mathrm{clip}}\}\}.
 \label{eq:clip-level}
\end{equation}
Then
\begin{equation}
 \left|
 \E Z_f^{\mathrm{ret}}
 -
 \E\operatorname{clip}_{B_{\mathrm{clip}}}
 (Z_f^{\mathrm{ret}})
 \right|
 \le
 \frac{\E|Z_f^{\mathrm{ret}}|^2}{B_{\mathrm{clip}}}
 \le\frac{\varepsilon}{32}.
 \label{eq:clip-bias}
\end{equation}
Run the combined sampler for $(N,\vct S)$ at total-variation distance
$d_{\mathrm{TV}}\le\varepsilon/[128(B_{\mathrm{clip}}+1)]$ from its target
law. Since the
clipped output has magnitude at most $B_{\mathrm{clip}}$, this changes its
expectation by at most $\varepsilon/64$. Its second moment under the
approximate law is at most
\begin{equation}
 \Sigma_{m,K}
 +2B_{\mathrm{clip}}^2d_{\mathrm{TV}}
 \le\frac32\Sigma_{m,K},
 \label{eq:approximate-second-moment}
\end{equation}
so the same polynomial sample bound applies. Because clipping is
$1$-Lipschitz, evaluating $Z_f^{\mathrm{ret}}$ to error
$\varepsilon/64$ per sample changes the final mean by at most the same
amount. For a target failure probability $0<\delta<1$, median-of-means then
requires
\begin{equation}
 O\!\left(
 \frac{\Sigma_{m,K}}{\varepsilon^2}
 \log\frac1\delta
 \right)
 \label{eq:MOM-sample-count}
\end{equation}
samples to attain additive statistical error at most $\varepsilon/4$ with
failure probability
at most $\delta$. For fixed $K$ and inverse-polynomial
$\varepsilon,\delta$, both the sample count and the bit cost per sample are
polynomial in $m$. Combining the physical tail, rare-sector,
correlation-truncation, sampling, clipping, arithmetic, and statistical
errors proves Theorem~\ref{thm:bounded-functions} in the standard bit model.

\subsection{Extension to general outcome functions}

Section~\ref{sec:general-functions} invokes the bounded algorithm only for
the functions $f_{m,j}^{(\mathrm c)}$ and $f_{m,j}^{(\mathrm s)}$ in
Eq.~\eqref{eq:bounded-characteristic-scores}. To evaluate either one to
absolute error at most $\zeta$, request an approximation to $f_m(\vct s)$
with error at most
\begin{equation}
 \Delta_j
 :=\min\!\left\{1,\frac{\zeta T_j}{8\pi}\right\},
\end{equation}
reduce that polynomial-bit number modulo $T_j$, and evaluate the sine or
cosine to error at most $\zeta/2$. If $\Delta_j<1$, the induced phase error
is at most $\zeta/4$; if $\Delta_j=1$, the definition of $\Delta_j$ gives
the same bound. The Lipschitz bound therefore gives total error below
$\zeta$. Since $\log T_j=\poly(m)$ and $f_m$ has polynomial-bit output, the
modular reduction and trigonometric evaluation take polynomial time.

Choose the internal bounded-estimation accuracy so that the total errors in
$\widehat c_j$ and $\widehat s_j$ are at most the $\eta$ in
Eq.~\eqref{eq:phase-scales}. Equation~\eqref{eq:characteristic-total-error}
then keeps $|\widehat z_j|$ bounded away from zero, allowing its principal
argument to be evaluated to the required polynomial-bit precision. Moreover,
$R,T_j$, and the integers in Eq.~\eqref{eq:dyadic-unwrapping} have
polynomial bit length, while Lemma~\ref{lem:phase-unwrapping} separates the
candidate lifts by a constant fraction of $T_j$.  After a constant-factor
tightening of the internal budgets, it suffices to evaluate every residue
representative to circular error at most $\varepsilon/4096$ modulo $T_j$ and
every lifted value to absolute error at most $\varepsilon/4096$.  This
requires only $O(\log T_J+\log\varepsilon^{-1})=\poly(m)$ bits, is far below
the lift separation, and contributes only a fixed fraction of the final
error budget. This completes the
bit-complexity proof of Theorem~\ref{thm:main-supp}.

\section{Uniform local coefficient bound for the tilted pair law}
\label{app:local-CLT}

This appendix proves Lemma~\ref{lem:saddle-extraction}.  Put $s=s_N$,
$y_j=sx_j$, and define
\begin{equation}
 \varphi(\theta)
 :=\frac{\mathcal Z(se^{i\theta})}{\mathcal Z(s)}
 =\prod_{j=1}^m
 \left(\frac{1-y_j}{1-y_je^{i\theta}}\right)^{1/2}.
 \label{eq:characteristic-function}
\end{equation}
This is the characteristic function of the tilted sum $Q=\sum_jQ_j$.
Fourier inversion gives
\begin{equation}
 p_N(s)=\frac1{2\pi}\int_{-\pi}^{\pi}
 e^{-iN\theta}\varphi(\theta)\dd\theta.
 \label{eq:Fourier-inversion}
\end{equation}

For the $L^2$ bound, observe that
\begin{equation}
 \log|\varphi(\theta)|^2
 =-\frac12\sum_j
 \log\!\left[
 1+\frac{2y_j(1-\cos\theta)}{(1-y_j)^2}
 \right].
 \label{eq:phi-modulus}
\end{equation}
Under $y_j\le1-\delta_{\rm fg}$, the arguments added to one are uniformly
bounded.  Using $\log(1+u)\ge u/(1+u_{\max})$ and
Eq.~\eqref{eq:tilted-variance},
\begin{equation}
 |\varphi(\theta)|^2
 \le\exp[-c_\delta\sigma^2(1-\cos\theta)]
 \le\exp[-c'_\delta N\theta^2],
 \qquad |\theta|\le\pi.
 \label{eq:phi-Gaussian-decay}
\end{equation}
Consequently,
\begin{equation}
 \frac1{2\pi}\int_{-\pi}^{\pi}|\varphi(\theta)|^2\dd\theta
 \le\frac{C_\delta}{\sqrt{N+1}}.
 \label{eq:phi-L2}
\end{equation}

For the point-mass lower bound, center the characteristic function:
$\psi(\theta)=e^{-iN\theta}\varphi(\theta)$.  All cumulants of order
$r\ge2$ are sums of the single-mode cumulants.  Uniform differentiation of
$-\frac12\log(1-ye^t)$ for $y\le1-\delta_{\rm fg}$ gives
\begin{equation}
 |\kappa_r(Q)|\le C_{r,\delta}N.
 \label{eq:cumulant-bound}
\end{equation}
Hence, for sufficiently small fixed $|\theta|$,
\begin{equation}
 \log\psi(\theta)
 =-\frac12\sigma^2\theta^2+R(\theta),
 \qquad |R(\theta)|\le C_\delta N|\theta|^3.
 \label{eq:centered-expansion}
\end{equation}
Choose a fixed $A_0$ large enough that the Gaussian-decay integral outside
$|\theta|\le A_0/\sqrt N$ is a small fraction of $N^{-1/2}$.  On this central
interval Eq.~\eqref{eq:centered-expansion} has uniformly small phase and
relative-modulus error for sufficiently large $N$, while
$N\le\sigma^2\le N/\delta_{\rm fg}$.  Therefore
\begin{equation}
 \Re\int_{-A_0/\sqrt N}^{A_0/\sqrt N}\psi(\theta)\dd\theta
 \ge\frac{c_\delta}{\sqrt N}.
 \label{eq:central-positive}
\end{equation}
Equation~\eqref{eq:phi-Gaussian-decay} bounds the complementary integral by
less than half this value.  Hence there exist
$N_0=N_0(\delta_{\rm fg})$ and $c_\delta'>0$ such that
\begin{equation}
 p_N(s_N)\ge\frac{c_\delta'}{\sqrt N},
 \qquad N\ge N_0.
 \label{eq:large-N-local-lower}
\end{equation}

It remains to make the bound uniform for
$1\le N<N_0$.  Under the tilted law,
\begin{equation}
 p_N(s_N)
 =\left(\prod_j\sqrt{1-y_j}\right)
 [z^N]\prod_j(1-y_jz)^{-1/2}.
 \label{eq:small-N-mass}
\end{equation}
Since $-\log(1-y)\le y/(1-y)$,
\begin{equation}
 -\log\prod_j\sqrt{1-y_j}
 \le\frac12\sum_j\frac{y_j}{1-y_j}
 =N,
 \label{eq:tilted-vacuum-lower}
\end{equation}
and therefore the prefactor in Eq.~\eqref{eq:small-N-mass} is at least
$e^{-N}$.  Moreover,
\begin{equation}
 \prod_j(1-y_jz)^{-1/2}
 =
 \exp\!\left[
 \sum_{\ell\ge1}\frac{z^\ell}{2\ell}
 \sum_jy_j^\ell
 \right],
 \label{eq:tilted-coefficient-exponential}
\end{equation}
whose coefficients are nonnegative.  Retaining only the contribution from
the linear term in the exponent gives
\begin{equation}
 [z^N]\prod_j(1-y_jz)^{-1/2}
 \ge\frac1{N!}
 \left(\frac12\sum_jy_j\right)^N.
 \label{eq:small-N-coefficient-lower}
\end{equation}
The fugacity gap and the saddle equation imply
\begin{equation}
 2N=\sum_j\frac{y_j}{1-y_j}
 \le\delta_{\rm fg}^{-1}\sum_jy_j,
\end{equation}
so that
\begin{equation}
 p_N(s_N)
 \ge e^{-N}\frac{(\delta_{\rm fg}N)^N}{N!}.
 \label{eq:small-N-uniform-lower}
\end{equation}
Because only the finitely many integers $1\le N<N_0$ remain, decreasing
$c_\delta'$ if necessary yields
\begin{equation}
 p_N(s_N)\ge\frac{c_\delta}{\sqrt{N+1}}
 \qquad\text{for every }N\ge1,
 \label{eq:all-N-local-lower}
\end{equation}
uniformly over all allowed squeezing profiles.  The vacuum sector $N=0$ is
handled separately: the auxiliary tilted law is then the point mass at
$Q=0$.  This should not be confused with the physical vacuum probability
$q_0=\prod_j\sqrt{1-x_j}$, which need not equal one.

Finally, Cauchy's formula on the saddle circle gives
\begin{equation}
 \frac{[z^N]\mathcal Z(z)F(z)}{\mathcal Z_N}
 =\frac1{2\pi p_N(s)}\int_{-\pi}^{\pi}
 e^{-iN\theta}\varphi(\theta)F(se^{i\theta})\dd\theta.
 \label{eq:coefficient-Fourier}
\end{equation}
Cauchy--Schwarz, Eq.~\eqref{eq:phi-L2}, and
$p_N(s)^{-2}=O_\delta(N)$ prove
Eq.~\eqref{eq:saddle-extraction}.

\section{Haar moments of quadratic and two-frame bilinear forms}
\label{app:Haar-moments}

For an integer $q\ge1$, let $u$ be uniform on the complex unit sphere and let
$B=B^T$ have operator norm $L$ and Hilbert--Schmidt norm squared $S$.  If a
standard complex Gaussian vector is written $g=Ru$, then $R$ and $u$ are
independent and
$\E R^{4q}=\Poch m{2q}$.  Takagi factorization and a Gaussian generating
function give
\begin{equation}
 \E|g^TBg|^{2q}
 =4^q(q!)^2[z^q]\det(\Id-zBB^\dagger)^{-1/2}.
 \label{eq:Gaussian-quadratic}
\end{equation}
If $b_j$ are the Takagi singular values, then, for every integer $r\ge1$,
$\sum_jb_j^{2r}\le SL^{2r-2}$, and therefore, coefficientwise for $z\ge0$,
\begin{equation}
 \det(\Id-zBB^\dagger)^{-1/2}
 \preccurlyeq(1-L^2z)^{-S/(2L^2)}.
 \label{eq:det-bound}
\end{equation}
Dividing by the radial moment yields
\begin{equation}
 \E|u^TBu|^{2q}
 \le\frac{4^qq!L^{2q}\Poch{S/(2L^2)}q}
 {\Poch m{2q}}.
 \label{eq:quadratic-moment-exact}
\end{equation}
The zero-operator case is understood by continuity.

Now let $(u,v)$ be a Haar orthonormal two-frame and let $C$ have operator
norm $L$ and Hilbert--Schmidt norm squared $S$.  Conditional on $u$, the
vector $v$ is uniform on the unit sphere of $u^\perp$, so
\begin{equation}
 \E_v[|u^TC\overline v|^{2q}\mid u]
 \le\frac{q!}{\Poch{m-1}q}\norm{C^Tu}^{2q}.
 \label{eq:conditional-v-moment}
\end{equation}
The complex-sphere moment formula for the positive operator
$(C^T)^\dagger C^T=\overline C\,C^T$ gives
\begin{equation}
 \E_u\norm{C^Tu}^{2q}
 \le\frac{L^{2q}\Poch{S/L^2}q}{\Poch mq}.
 \label{eq:Ctu-moment}
\end{equation}
Thus
\begin{equation}
 \E_{u\perp v}|u^TC\overline v|^{2q}
 \le\frac{q!L^{2q}\Poch{S/L^2}q}
 {\Poch mq\Poch{m-1}q}.
 \label{eq:bilinear-moment-exact}
\end{equation}
Using $q!\le q^q$, $\Poch{a}{q}\le(a+q)^q$, and the lower bounds on the
denominators, Eqs.~\eqref{eq:quadratic-moment-exact} and
\eqref{eq:bilinear-moment-exact} imply
\[
 \E|u^TBu|^{2q}
 \le\left[\frac{4q(S/2+qL^2)}{m^2}\right]^q,
 \qquad
 \E_{u\perp v}|u^TC\overline v|^{2q}
 \le\left[\frac{2q(S+qL^2)}{m^2}\right]^q
\]
for $m\ge2$.  On the saddle circle,
Eq.~\eqref{eq:contraction-norms} gives $S=O_\delta(N)$ and
$L=O_\delta(1)$; the moment orders used above satisfy $q\le3k\le3N$.
These observations prove Eq.~\eqref{eq:simple-Haar-moments}, including its
conjugated variants.

\section{Optimized linear-order envelope}
\label{app:Stirling-envelope}

We prove Corollary~\ref{cor:linear-orders}.  Write
\begin{equation}
 n=\rho m+O(1),\qquad
 k=\alpha m+O(1),\qquad
 \ell=\beta m+O(1),
 \label{eq:linear-scaling}
\end{equation}
where $0<\alpha\le\beta\le\rho<\infty$, and define
\begin{equation}
 h(x):=(1+x)\log(1+x)-x\log x,
 \qquad h(0)=0.
 \label{eq:h-entropy}
\end{equation}
Uniform Stirling estimates applied to Eq.~\eqref{eq:envelope-E} give
\begin{equation}
 \frac1m\log\mathcal E_{m,n}(k,\ell)
 =\Phi(\rho,\alpha,\beta)
 +O\!\left(\frac{\log m}{m}\right),
 \label{eq:Phi-asymptotic}
\end{equation}
where
\begin{align}
 \Phi(\rho,\alpha,\beta)
 ={}&h(\rho)-2h(\alpha)
 +2\rho\log\rho-2\beta\log\beta\notag\\
 &+(\beta-\alpha)\log(\beta-\alpha)
 -(\rho-\alpha)\log(\rho-\alpha)\notag\\
 &-(1+\rho+\alpha)\log(1+\rho+\alpha)
 +(1+\beta+\alpha)\log(1+\beta+\alpha).
 \label{eq:Phi}
\end{align}
The convention $0\log0=0$ covers boundaries.  Differentiation gives
\begin{equation}
 \partial_\beta\Phi
 =\log\frac{(\beta-\alpha)(1+\beta+\alpha)}{\beta^2},
 \label{eq:Phi-derivative}
\end{equation}
so the minimizer is
\begin{equation}
 \beta_\star=\min\{\rho,\alpha(1+\alpha)\}.
 \label{eq:beta-star}
\end{equation}
If the minimizer is internal, substitution gives
\begin{align}
 \Phi_\star(\rho,\alpha)
 ={}&(1+\rho)\log(1+\rho)+\rho\log\rho\notag\\
 &-(\rho-\alpha)\log(\rho-\alpha)
 -(1+\rho+\alpha)\log(1+\rho+\alpha).
 \label{eq:Phi-internal}
\end{align}
It vanishes at $\alpha=0$ and
\begin{equation}
 \partial_\alpha\Phi_\star
 =\log\frac{\rho-\alpha}{1+\rho+\alpha}<0.
 \label{eq:Phi-alpha}
\end{equation}
If the minimizer is the boundary $\beta=\rho$, then
$\Phi_\star=h(\rho)-2h(\alpha)$.  At the joining point this equals the
strictly negative internal value, and it decreases for larger $\alpha$
because $-2h'(\alpha)<0$.  Thus
\begin{equation}
 \min_{\alpha\le\beta\le\rho}
 \Phi(\rho,\alpha,\beta)<0
 \quad\text{for every }\alpha>0.
 \label{eq:negative-rate}
\end{equation}
On the compact set
$\vartheta\le\alpha\le\rho\le\rho_0$, the optimized rate is bounded above
by a negative constant.  The $O(\log m)$ Stirling remainder contributes only
a polynomial prefactor, proving Eq.~\eqref{eq:linear-order-bound}.

\section{Symbol index}
\label{app:symbol-index}

\begingroup
\setlength{\parindent}{0pt}
\setlength{\parskip}{2pt}

\(m\): number of optical modes.\par
\(N\), \(n=2N\): pair number and total photon number.\par
\(\mathcal H_{m,n}\), \(D_{m,n}\): fixed-\(n\)-photon Hilbert space and its dimension.\par
\(\Phi_{m,n}\), \(\nu_{m,n}\): fixed-total PNR patterns and their uniform law.\par
\(t_j\), \(x_j\): complex squeezing amplitude and \(\lvert t_j\rvert^2\).\par
\(\mathcal Z_N\), \(q_N\): sector normalization and physical pair-number probability.\par
\(\ket{\Omega_N}\), \(\hat\rho_{2N}\): normalized fixed-\(2N\)-photon input state and density operator.\par
\(\hat U_n\), \(\mathscr C_n(U)\): fixed-sector interferometer and its conjugation action.\par
\(\mathcal O_{n,\le k}\): space generated by number-preserving monomials with at most \(k\) creation--annihilation pairs.\par
\(\mathcal V_{n,k}\), \(\mathcal Q_{n,k}\): correlation-order module and its orthogonal projector.\par
\(d_{m,k}\): dimension of \(\mathcal V_{n,k}\).\par
\(\mathsf H_{n,k}\): projector onto the order-\(k\) multivariate Hahn subspace.\par
\(f_m\), \(f_{n,k}\): outcome function (written as \(f\) in the main text) and its order-\(k\) Hahn component in the \(n\)-photon sector.\par
\(\hat O_{f,n,k}\): PNR-diagonal operator induced by \(f_{n,k}\).\par
\(\mu_f(U)\), \(\widetilde\mu_f(U)\): exact outcome-function mean and its classical estimate.\par
\(\mu_{f,n}^{(k)}(U)\): correlation-order-\(k\) contribution in the \(n\)-photon sector.\par
\(\hat\gamma_N^{(k)}\): normalized \(k\)-particle reduced state.\par
\(\Lambda_{\rho,n,k}\): dimension-normalized state weight at correlation order \(k\).\par
\(w_{N,k}(\vct s;U)\), \(W_{N,K}(\vct s;U)\): exact-order and retained diagonal weights for a complete pattern.\par
\(K\), \(L_m(B)\), \(B\): correlation-order cutoff, pair-number cutoff, and physical-tail exponent.\par
\(M_m\), \(V_m\): known mean envelope and physical output-variance bound for the general function class.\par
\(R\), \(T_j\), \(J\), \(\eta\): local phase scale, dyadic periods, number of scales, and bounded-estimation accuracy.\par
\(\Sigma_{m,K}\): polynomial second-moment bound for the bounded signed-weight estimator.\par

\endgroup

\bibliography{references}